\documentstyle[12pt]{article}

\begin{document}
\input{psfig.sty}
\begin{flushright}
\baselineskip=12pt
\end{flushright}

\begin{center}
\vglue 1.5cm
{\Large\bf Gauge-Fermion Unification and Flavour Symmetry}
\vglue 2.0cm
{\Large Tianjun Li\footnote{
E-mail: tli@sns.ias.edu. Phone Number: (609) 734-8024. Fax Number:  (609) 951-4489.}}
\vglue 1cm
{School of Natural Science, Institute for Advanced Study,  \\
             Einstein Drive, Princeton, NJ 08540, USA\\  }
\end{center}

\vglue 1.5cm
\begin{abstract}
After we study the 6-dimensional ${\cal N} = (1, 1)$
supersymmetry breaking and $R$ symmetry breaking on $M^4\times T^2/Z_n$,
we construct two ${\cal N} = (1, 1)$ supersymmetric $E_6$ models
 on $M^4\times T^2/Z_3$ where
$E_6$ is broken down to $SO(10)\times U(1)_X$ by orbifold projection.
In Model I, three families of the Standard Model fermions arise
from the zero modes of bulk vector multiplet, and the 
$R$ symmetry $U(1)_F^{I} \times SU(2)_{{\bf 4}_-}$
can be considered as flavour symmetry. This may explain why
there are three families of fermions in the nature. 
In Model II, the first two families come
from the zero modes of bulk vector multiplet, and 
the flavour symmetry is similar.
In these models, the anomalies can be cancelled, and
 we have very good fits to 
the SM fermion masses and mixings.
We also comment on the
${\cal N}=(1, 1)$ supersymmetric $E_6$ models on
 $M^4\times T^2/Z_4$ and $M^4\times T^2/Z_6$, $SU(9)$ models on $M^4\times T^2/Z_3$,
and $SU(8)$ models on $T^2$ orbifolds.
\\[1ex]
\\[1ex]

\end{abstract}

\vspace{0.5cm}
\begin{flushleft}
\baselineskip=12pt
\today \\
\end{flushleft}
\newpage
\setcounter{page}{1}
\pagestyle{plain}
\baselineskip=14pt

\section{Introduction}
Supersymmetric Grand Unified Theory (SUSY GUT) is one of the most
promising candidates which describe the fundamental
interactions except gravity in the nature.
Supersymmetry is an elegant solution to the
gauge hierarchy problem, and the Grand Unified Theory gives us a simple
 understanding of the quantum numbers of quarks and leptons. Moreover,
the success of gauge coupling unification
 in the Minimal Supersymmetric Standard Model strongly supports
 this idea. The electroweak symmetry can be broken by
radiative corrections due to the large top quark Yukawa coupling,
 and the very tiny neutrino masses 
can be realized naturally by see-saw mechanism in the SUSY GUT.

Among several candidates of the GUT gauge groups such as SU(5), SO(10),
and E$_6$, SO(10) has several particularly attractive features. 
 SO(10) is the smallest semi-simple GUT group
which does not require a particular particle content to cancel the
gauge anomaly. And there exists
the fermion unification, {\it i.e.},  all the quarks and leptons including 
the right-handed neutrinos in each generation are unified to a single {\bf 16}
dimensional spinor representation field, and then, we have
the Yukawa unification.
In addition, the doublet-triplet splitting can be explained by
the Dimopoulos-Wilczek mechanism~\cite{DWDT}.

However, it is very difficult to understand the fermion masses and mixings
in SO(10) model due to the fermion unification. The Yukawa matrices
have hierarchical structures and they are quite different among
the quark and lepton sectors.
On the one hand, all the mixing angles in the quark sector are small
and the masses of quarks have hierarchical structures. The magnitude of 
the mass hierarchy is enormous, for instance, 
the ratio of the up-quark mass $m_u$ to the top-quark
mass $m_t$ is approximately $10^{-5}$. On the other hand, there exist
the bilarge mixings in neutrino-lepton sector from solar neutrino, 
atmospheric neutrino 
and reactor neutrino experiments~\cite{sno, skatm, kamland}.
$\theta_{12}$, which is the mixing angle for solar oscillation, is
large with best fit around $32.6^o$.
 And $\theta_{23}$, which is the mixing angle for atmospheric oscillation,
is also large with best fit around $45^o$~\cite{phays, VBDM}. 
However, the mixing angle $\theta_{13}$ is small, 
$\sin^2\theta_{13} \leq 0.04$ from the CHOOZ experiment~\cite{chooz}. 

How to understand the fermion masses and mixings in $SO(10)$ model is an
interesting question. The Froggatt-Nielsen (FN) mechanism is an elegant scenario
to generate the hierarchical structures in Yukawa matrices~\cite{FN}. 
Various flavour
symmetries have been introduced to explain the fermion masses and mixings
 in the supersymmetric $SO(10)$ models, for example,
$U(1)$ and extra discrete flavour symmetry~\cite{CASB, KBWP},
 $U(2)$ flavour symmetry~\cite{FSU2},
$SU(2)\times U(1)$ flavour symmetry~\cite{SRABY1, SRABY2}, 
$SU(2)\times Z_2 \times Z_2\times Z_2$ flavour symmetry~\cite{MCKTM},
$SU(3)\times U(1)$ flavour symmetry~\cite{RKYM}, 
$SU(3)$ flavour symmetry~\cite{FSSU3}, etc.
And the fermion-mass hierarchy in 5-dimensional 
SO(10) models can be realized by different fermions with different
 5-dimensional wave-function profiles.  Unlike the Froggatt-Nielsen mechanism, 
the SO(10) breaking effect may directly contribute to the source of the fermion
mass hierarchy~\cite{RKTL}.
Moreover, if we introduce extra {\bf 10}-dimensional
matter multiplets and mix or flip the $d^c$ and doublet $L$ in
 the {\bf 16} and those in the {\bf 10}-dimensional matter multiplets 
by the SO(10) breaking effect,
we can explain the fermion masses and mixings by introducing $U(1)$
flavour symmetry~\cite{QSZT1, NMU1}. 
This approach can be generalized to the 5-dimensional
$SO(10)$ models~\cite{QSZT2}, and to the
 6-dimensional $SO(10)$ models where the Yukawa couplings can also be suppressed
by the volume suppression factors~\cite{HABA}.
 However, this approach loses the fermion 
unification. By the way, the fermion masses and mixings in the $SO(10)$ models
with various flavour symmetries has been reviewed in Ref.~\cite{MCKMrev}.

In addition, a new kind of scenario for the GUT breakings has
 been discussed extensively during the last several 
years~\cite{SBPL,JUNI, HJLL, JUNII, TWTY, GMSN}.
 The key point is that the supersymmetric GUT
models exist in 5 or higher dimensions and are broken down to the
4-dimensional ${\cal N}=1$ supersymmetric Standard like Models for
 the zero modes due to the discrete symmetries on the extra space manifolds, which
become non-trivial constraints on the multiplets and gauge generators in the
GUTs~\cite{JUNII}. The attractive models have been constructed explicitly, where
the supersymmetric 5-dimensional and 6-dimensional 
GUT models are broken down to the 4-dimensional ${\cal N}=1$
 supersymmetric $SU(3)\times SU(2) \times U(1)^{n-3}$
models, where $n$ is the rank of GUT group, through the 
compactification on various orbifolds. 
The GUT gauge symmetry breaking and doublet-triplet
splitting problems have been solved neatly by the
discrete symmetry projections. And the partial gauge-fermion unification
and the family symmetry as a remnant of the higher-dimensional $R$
symmetry have been discussed in the
6-dimensional  ${\cal N} = (1,1)$ supersymmetric $SO(10)$
model on the space-time $M^4\times T^2/Z_3$
where $M^4$ is the 4-dimensional Minkowski space-time~\cite{TWTY}. 
Other interesting phenomenology, like $\mu$ problem, gauge coupling
unification, non-supersymmetric GUT, gauge-Higgs unification,
gauge-Yukawa unification, 
proton decay, etc, have also been 
studied~\cite{SBPL,JUNI, HJLL, JUNII, TWTY, GMSN}. 

Before we present our results, 
let us ask three interesting questions\\

(1) The maximal supersymmetry in the supersymmetric gauge theory is 
the 4-dimensional ${\cal N}=4$ supersymmetry, why is not the 4-dimensional 
${\cal N}=4$ supersymmetry in the nature?\\

(2) Why are there three families of fermions 
in the Standard Model (SM)~\cite{Adler}?\\

(3) If $SO(10)$ model is correct, what is the origin of the flavour symmetry
which is used to explain the fermion masses and mixings?\\

In this paper, we first briefly review the $T^2/Z_n$ orbifolds. And then, 
we consider the 6-dimensional ${\cal N} = (1, 1)$ supersymmetric gauge theory
on the space-time $M^4\times T^2$, which corresponds to the 4-dimensional
${\cal N}= 4$ supersymmetry, the maximal supersymmetry
for gauge theory. The $R$ symmetry in this case is
$SO(2)_{56}\times SU(2)_{\bf 4_+}\times SU(2)_{\bf 4_-}$, and 
the transformation properties of vector multiplet under
this $R$ symmetry are given. 
In order to preserve only the 4-dimensional ${\cal N} = 1$ 
supersymmetry, we consider the supersymmetry breaking and $R$ symmetry breaking
on $M^4\times T^2/Z_n$. We find
 that for $T^2/Z_3$ orbifold, the
$R$ symmetry is $SO(2)_{56}\times U(1)_{{\bf 4}_+}
\times SU(2)_{{\bf 4}_-}$. And for $T^2/Z_4$ and $T^2/Z_6$ orbifolds, 
 the $R$ symmetry is $SO(2)_{56}\times U(1)_{{\bf 4}_+} 
\times  U(1)_{{\bf 4}_-}$.
We also present the transformation properties of the vector multiplet under
these $R$ symmetries.
In addition, we explain how to embed the $R$ symmetry into the $SU(4)_R$ $R$
symmetry, which is the maximal $R$ symmetry for
the 4-dimensional ${\cal N} = 4$ supersymmetry. This embedding gives us
 an elegant way to understand the supersymmetry and $R$ symmetry breakings.

In the 6-dimensional ${\cal N} = (1, 1)$ supersymmetric gauge theory, there
is a vector multiplet, which in the 4-dimensional ${\cal N} = 1$ supersymmetry language
corresponds to one gauge supermultiplet and three chiral multiplets in
the adjoint representation. The Standard Model fermions may come from
 three chiral multiplets due to the orbifold gauge symmetry breaking, which
may explain why there are three families of fermions in the 
Standard Model. And the $R$ symmetry can give us the flavour symmetry to
explain the fermion masses and mixings. 
To be concrete, we discuss the
${\cal N} = (1, 1)$ supersymmetric $E_6$ models
 on the space-time $M^4\times T^2/Z_3$ where the
$E_6$ gauge symmetry
is broken down to $SO(10)\times U(1)_X$ by orbifold projection. 
There are three $SO(10)$ spinor representations ${\bf 16}$ from the
zero modes of bulk vector multiplet, which can be considered
as three families of fermions. We study two models in detail.
In Model I, three families of the Standard Model fermions arise
from the zero modes of bulk vector multiplet, and the 
$R$ symmetry $U(1)_F^{I} \times SU(2)_{{\bf 4}_-}$
can be considered as flavour symmetry. Moreover,
we show that the anomalies can be cancelled, and 
the very good fits to 
the Standard Model fermion (quark, lepton, and neutrino) masses and mixings 
can be obtained. In this model, we can understand why there are
three families of the SM fermions in the nature and the origin of flavour symmetry.
 However, the $U(1)_R^{I}$ $R$ symmetry is not
the traditional $R$ symmetry in the low energy supersymmetry phenomenology
and is broken by the Higgs VEVs. Because 
 there is no unbroken 
$Z_2$ symmetry under which the ordinary particles are even while their superpartners
are odd, there are no cold dark matter candidates (for example neutralino)
from the superpartners. We point out that
this $U(1)_R^I$ $R$ symmetry may be the unbroken $U(1)_R$ R symmetry 
in the compactification of the weakly coupled
heterotic string theory on Calabi-Yau manifolds. So, it is interesting to
study its low energy phenomenology.
In Model II, the first two families of the Standard Model fermions arise
from the zero modes of bulk vector multiplet, and the 
$R$ symmetry $U(1)_F^{II} \times SU(2)_{{\bf 4}_-}$
can be considered as flavour symmetry. We also discuss
the anomaly cancellations, and the fermion masses and mixings.
Similar to those in Model I, we have very good fits to 
the Standard Model fermion masses and mixings. 
Unlike that in Model I, the $U(1)_R^{II}$ $R$ symmetry is the
traditional $U(1)_R$ 
$R$ symmetry in the low energy supersymmetry phenomenology.
However, we can not explain why there are
three families of fermions in the nature.

Furthermore, we comment on the
${\cal N}=(1, 1)$ supersymmetric $E_6$ models on
 $M^4\times T^2/Z_4$ and $M^4\times T^2/Z_6$ where there
are two families of the Standard Model fermions from the zero modes
of bulk vector multiplet. In these models, 
the $U(1)_R^{III}$ $R$ symmetry is the traditional $U(1)_R$ $R$ symmetry in the
low energy supersymmetry phenomenology, and
the $U(1)_F^{III}\times U(1)_{RF}$ $R$ symmetry can be considered
as flavour symmetry.  In addition, we comment on the
${\cal N}=(1, 1)$ supersymmetric $SU(9)$ and $SU(8)$
models on $M^4\times T^2/Z_3$ where there are three or two families of 
the SM fermions
from the zero modes of bulk vector multiplet, and comment
on the $SU(8)$ models on $M^4\times T^2/Z_4$ and $M^4\times T^2/Z_6$.
The $R$ symmetry and flavour symmetry are similar to those in $E_6$ models
on the corresponding $T^2$ orbifolds.

This paper is organized as follows: in Section 2, we briefly review the 
$T^2$ orbifolds, explain the 6-dimensional ${\cal N} = (1, 1)$ supersymmetric
gauge theory, and consider the supersymmetry breaking and $R$ symmetry
breaking by orbifold projection.
In Section 3, we discuss the ${\cal N}=(1, 1)$ supersymmetric $E_6$ models on
the space-time $M^4\times T^2/Z_3$. We comment on the
${\cal N}=(1, 1)$ supersymmetric $E_6$ models on
 $M^4\times T^2/Z_4$ and $M^4\times T^2/Z_6$ in Section 4. We also comment on the
${\cal N}=(1, 1)$ supersymmetric $SU(9)$ models on $M^4\times T^2/Z_3$
 and $SU(8)$ models on $T^2$ orbifolds in
Section 5.
The discussions and conclusions are given in Section 6.

\section{6-Dimensional ${\cal N}=(1, 1)$ Supersymmetric Gauge Theory on
 $M^4\times T^2/Z_n$}

\subsection{Discrete Symmetry on $T^2$ and $T^2$ Orbifolds}
 We consider the 6-dimensional space-time which can be factorized into a product of the 
ordinary 4-dimensional Minkowski space-time $M^4$, and the torus $T^2$
which is homeomorphic to $S^1\times S^1$. The corresponding
coordinates for the space-time are $x^{\mu}$, ($\mu = 0, 1, 2, 3$),
$ x^5$ and $ x^6$. 
The radii for the circles along $x^5 $ direction and $x^6 $ direction are
$R_1$ and $R_2$, respectively. We define the
complex coordinate
\begin{eqnarray}
z \equiv{1\over 2} \left(x^5 + i x^6\right)~.~\,
\end{eqnarray}
In the complex coordinate,
the torus $T^2$ can be defined by $C^1$ moduloed the
equivalent classes: 
\begin{eqnarray}
z \sim z+ \pi R_1 ~,~ z \sim z +  \pi R_2 e^{{\rm i}\theta} ~.~\,
\end{eqnarray}

The complete discrete symmetries on the
torus are $Z_2$, $Z_3$, $Z_4$ and $Z_6$~\cite{JUNI}. And
for $Z_3$, $Z_4$ and $Z_6$ discrete symmetries, we have to 
choose $R_1=R_2=R$.

For $Z_3$ symmetry, $\theta= 2\pi/3$. And
the equivalent class $z \sim z +  \pi R  e^{{\rm i}2\pi/3}$ 
is equivalent to the equivalent class
$z \sim z +  \pi R e^{{\rm i} \pi /3}$. 
This torus admits the $Z_3$ symmetry generated by
\begin{eqnarray}
\gamma_3~:~ z \rightarrow e^{{\rm i}2\pi/3}  z~.~
\end{eqnarray}
There are
three fixed points: $z=0$, $z=\pi R e^{{\rm i}\pi/6}/{\sqrt 3}$, and
$z= 2 \pi R e^{{\rm i}\pi/6}/{\sqrt 3}$.
The 3-branes can be located at the fixed points where the extra particles
can be put on.

For $Z_4$ symmetry, $\theta= \pi/2$. The $Z_4$ symmetry is generated by
\begin{eqnarray}
\gamma_4~:~ z \rightarrow e^{{\rm i}\pi/2}  z~.~
\end{eqnarray}
There are
two $Z_4$ fixed points: $z=0$ and $z={\sqrt 2} \pi R e^{{\rm i}\pi/4}/2$, and
two $Z_2$ fixed points: $z=\pi R/2$ and $z= \pi R e^{{\rm i}\pi/2}/2$.

For $Z_6$ symmetry, $\theta= \pi/3$. The $Z_6$ symmetry is generated by
\begin{eqnarray}
\gamma_4~:~ z \rightarrow e^{{\rm i}\pi/3}  z~.~
\end{eqnarray}
There is one $Z_6$ fixed point
$z=0$, two $Z_3$ fixed points: 
 $z=\pi R e^{{\rm i}\pi/6}/{\sqrt 3}$ and
$z=2 \pi R e^{{\rm i}\pi/6}/{\sqrt 3}$, and three $Z_2$ fixed points:
$z=\sqrt 3 \pi R e^{{\rm i}\pi/6}/2$, $z=\pi R/2$ and $z= \pi R e^{{\rm i}\pi/3}/2$.

In short, the $T^2/Z_n$ orbifolds are obtained from $T^2$ moduloed the
discrete symmetry $Z_n$. The KK mode expansions have been discussed in Ref.~\cite{JUNI}.

\subsection{6-Dimensional ${\cal N}=(1, 1)$ Supersymmetric 
Gauge Theory on $M^4\times T^2$}

As we know, the 6-dimensional
 ${\cal N}=1$ ((0, 1) or (1, 0)) supersymmetric gauge theory is
chiral, where the gaugino (and gravitino) 
and the matters (hypermultiplets) have opposite chiralities, so, 
it has anomalies which are usually difficult to be cancelled by introducing the
hypermultiplets. Therefore, we consider
the 6-dimensional ${\cal N}=(1, 1)$ supersymmetry, where the irreducible box anomalies 
and reducible anomalies vanish, and there is no global anomaly.

For the 6-dimensional ${\cal N}=(1, 1)$ supersymmetry,
there are four supercharges ${\cal Q}^{(6)}_{{\bf 4_+}, A}$ 
and ${\cal Q}^{(6)}_{{\bf 4_-}, B}$ with $A, B=1, 2$, which belong to
the ${\bf 4_+}$ and ${\bf 4_-}$ spinor representations of the $SO(5,1)$
with positive and negative chiralities, respectively~\cite{JS, YITWTY}.
The pseudo-Majorana condition is imposed on these four supercharges,
\begin{eqnarray}
 {\cal Q}_{{\bf 4}_{+}, A}^{(6)}~=~J_{A A'} ~ ({\cal Q}_{{\bf 4}_{+}, A'}^{(6)})^c
~,~  {\cal Q}_{{\bf 4}_{-}, B}^{(6)} ~=~J_{B B'} ~({\cal Q}_{{\bf 4}_{-}, B'}^{(6)})^c~,~\,
\end{eqnarray}
where $c$ is the charge conjugation, and $J$ is a symplectic invariant metric
\begin{eqnarray}
J=\left(\begin{array}{cc}
			0   & - 1 \\
              1 & 0
				\end{array}\right) ~.~\,
\end{eqnarray}
Thus, only two supercharges are independent.
The $R$ symmetry for the 6-dimensional ${\cal N}=(1, 1)$ supersymmetry is 
$SU(2)_{\bf 4_+}\times SU(2)_{\bf 4_-}$. The supercharges 
${\cal Q}^{(6)}_{{\bf 4_+}, A}$ and ${\cal Q}^{(6)}_{{\bf 4_-}, B}$
form the fundamental representations under the $R$ symmetry
$SU(2)_{\bf 4_+}$ and $SU(2)_{\bf 4_-}$, respectively~\cite{JS, YITWTY}.

For the gamma matrices and spinors
in 4-dimension and 6-dimension, we follow the convention in Ref.~\cite{YITWTY}.  
After the compactification on torus, each 6-dimensional supercharge is
decomposed into two 4-dimensional Weyl spinors
\begin{eqnarray}
 {\cal Q}_{{\bf 4}_{+}, 1}^{(6)} = \left(\begin{array}{cc}
		{\cal Q}^{(4)}_{\alpha, 1}& 0 \\
     0  & \bar{{\cal Q}}^{\dot{\alpha}, 2}_{(4)}
				      \end{array}\right)
\quad ~,~ \quad
 {\cal Q}_{{\bf 4}_{+}, 2}^{(6)} = \left(\begin{array}{cc}
		{\cal Q}^{(4)}_{\alpha, 2}& 0 \\
     0  & - \bar{{\cal Q}}^{\dot{\alpha}, 1}_{(4)}
				      \end{array}\right)~,~\,
\end{eqnarray}
\begin{eqnarray}
 {\cal Q}_{{\bf 4}_{-}, 1}^{(6)} = \left(\begin{array}{cc}
			0   &  {\cal Q}_{\alpha, 3}^{(4)} \\
              \bar{{\cal Q}}^{\dot{\alpha}, 4}_{(4)} & 0
				\end{array}\right)
\quad ~,~ \quad
 {\cal Q}_{{\bf 4}_{-}, 2}^{(6)} = \left(\begin{array}{cc}
			0   &  {\cal Q}_{\alpha, 4}^{(4)} \\
              -\bar{{\cal Q}}^{\dot{\alpha}, 3}_{(4)} & 0
				\end{array}\right)~,~\,
\end{eqnarray}
where ${\cal Q}_{\alpha, 1}^{(4)}$, ${\cal Q}_{\alpha, 2}^{(4)}$,
${\cal Q}_{\alpha, 3}^{(4)}$, and ${\cal Q}_{\alpha, 4}^{(4)}$
are the usual 4-dimensional supercharges that belong
to the $({\bf 1/2, 0})$ or ${\bf 2}$ spinor representation of the $SO(3,1)$,
and their charge conjugations $\bar{{\cal Q}}^{\dot{\alpha}, 1}_{(4)}$,
$\bar{{\cal Q}}^{\dot{\alpha}, 2}_{(4)}$, $\bar{{\cal Q}}^{\dot{\alpha}, 3}_{(4)}$,
and $\bar{{\cal Q}}^{\dot{\alpha}, 4}_{(4)}$ belong to the 
$({\bf 0,1/2})$ or ${\bf 2'}$ spinor representation of the $SO(3,1)$.
So, we have the 4-dimensional ${\cal N}=4$ supersymmetry. 
After the compactification, the $SO(2)_{56}$ rotation on the 
$x_5$-$x_6$ plane, which is a subgroup of the $SO(5,1)$, becomes the
 $R$ symmetry in the 4-dimensional supersymmetric gauge theory. Therefore, the
$R$ symmetry in the 4-dimensional supersymmetric gauge theory is 
$SO(2)_{56}\times SU(2)_{\bf 4_+}\times SU(2)_{\bf 4_-}$, which
is the subgroup of $SU(4)_R$, the maximal $R$ symmetry in 
the 4-dimensional ${\cal N}=4$ supersymmetric gauge theory. 
$({\cal Q}_{\alpha, 1}^{(4)}, ~{\cal Q}_{\alpha, 2}^{(4)})^T$
form a fundamental representation of $SU(2)_{\bf 4_+}$,
and $({\cal Q}_{\alpha, 3}^{(4)}, ~{\cal Q}_{\alpha, 4}^{(4)})^T$
form a fundamental representation of $SU(2)_{\bf 4_-}$ where
$T$ is the transpose. Moreover, since the supercharges
transform under the $SO(5,1)$ as spinor representations, they are 
charged under the $SO(2)_{56}$. From the 6-dimensional 
chiralities, we obtain that ${\cal Q}_{\alpha, 1}^{(4)}$ and
 ${\cal Q}_{\alpha, 2}^{(4)}$ carry the $SO(2)_{56}$ charges $+1/2$,
and ${\cal Q}_{\alpha, 3}^{(4)}$ and ${\cal Q}_{\alpha, 4}^{(4)}$
carry  charges $-1/2$.

In the 6-dimensional ${\cal N}=(1, 1)$ supersymmetric gauge theory, 
the vector multiplet is a unique supermultiplet of the $(1,1)$ supersymmetric 
theory besides the gravity multiplet.
It consists of a 6-dimensional vector field $A_\mu$ ($\mu=0,1, 2, 3$),
$A_5$, $A_6$,
two pseudo-Majorana-Weyl spinors with opposite chiralities
$\lambda_{{\bf 4_-}, A}$ and $\lambda_{{\bf 4_+}, B}$ where $A, B=1, 2$, and 
two complex scalar fields $\sigma^2$ and $\sigma^3$. All of
them are in the adjoint representation under the bulk gauge
group. The vector fields $A_\mu$, $A_5$, and $A_6$ are
invariant under the R-symmetry $SU(2)_{\bf 4_+} \times SU(2)_{\bf 4_-}$.
The spinors $(\lambda_{{\bf 4_-}, 1}, \lambda_{{\bf 4_-}, 2})^T$ 
transform as a fundamental representation under 
the $SU(2)_{\bf 4_+}$, and the spinors 
$(\lambda_{{\bf 4_+}, 1}, \lambda_{{\bf 4_+}, 2})^T$ transform as a
fundamental representation under
 the $SU(2)_{\bf 4_-}$. Four real components of scalar fields $\sigma^2$ and $\sigma^3$
belong to the $({\bf {\bar 2}}, ~{\bf {\bar 2}})$ representation 
under R-symmetry $SU(2)_{\bf 4_+} \times SU(2)_{\bf 4_-}$, which 
transforms as
\begin{eqnarray}
\Delta_{AB} \longrightarrow  [SU(2)_{{\bf 4_+}}]_A^{~A'}~ \Delta_{A'B'}~
[SU(2)_{{\bf 4}_-}]_B^{~B'}~,~\,
\end{eqnarray}
where
\begin{eqnarray}
\Delta_{AB} ~=~\epsilon_{A' A} \left(\begin{array}{cc}
 \sigma^2 & \sigma^3 \\  \sigma^{3*} & - \sigma^{2*}
      \end{array}
\right)^{A'B'} \epsilon_{B' B} ~.~\,
\end{eqnarray}
So, they form a vector of the $SO(4)_{\rm R} \simeq
SU(2)_{\bf 4_+}\times SU(2)_{\bf 4_-}$.

If the 6-dimensional
${\cal N}=(1,1)$ supersymmetric gauge theory is compactified
 on a torus ${\bf T}^2$, 
we have 4-dimensional ${\cal N}=4$ supersymmetry due to  the flat metric
of torus. The 4-dimensional
${\cal N}=4$ vector multiplet contains a 4-dimensional 
vector field $A_{\mu}$, four fermion partners 
$\chi_{\alpha}^a$ ($a=1,...,4$), and three complex scalar 
fields $\sigma^1, \sigma^2, \sigma^3$.
The four Weyl fermions $\chi_{\alpha}^a$ are obtained from the
6-dimensional fermion fields $\lambda_{\bf 4_-, A}$ and $\lambda_{\bf 4_+, B}$ 
as the following decomposition
\begin{equation}
\lambda_{{\bf 4_-}, 1} \rightarrow
\left(\begin{array}{cc}
0 & - \chi_{\alpha}^2  \\
  \bar{\chi}^{\dot{\alpha}}_1 & 0 
      \end{array}\right) \quad ~,~ \quad
\lambda_{{\bf 4_-}, 2} \rightarrow
\left(\begin{array}{cc}
0 &  \chi_{\alpha}^1  \\
  \bar{\chi}^{\dot{\alpha}}_2 & 0 
      \end{array}\right)~,~\, 
\end{equation}
\begin{equation}
\lambda_{{\bf 4_+}, 1} \rightarrow 
\left( \begin{array}{cc}
  \chi_{\alpha}^4 & 0 \\ 
0 & - \bar{\chi}^{\dot{\alpha}}_3
	\end{array}\right)
\quad ~,~ \quad
\lambda_{{\bf 4_+}, 2} \rightarrow 
\left( \begin{array}{cc}
 - \chi_{\alpha}^3 & 0 \\ 
0 & - \bar{\chi}^{\dot{\alpha}}_4
	\end{array}\right)~,~\, 
\end{equation}
where the $(\chi_{\alpha}^1, \chi_{\alpha}^2)^T$ and 
$(\chi_{\alpha}^3, \chi_{\alpha}^4)^T$ transform as 
the anti-fundamental representations under 
the $SU(2)_{\bf 4_+}$ and $SU(2)_{\bf 4_-}$, respectively.
And the complex field $\sigma_1$ is obtained from the
gauge fields $A_5$ and $A_6$ 
\begin{equation}
\sigma^1(x^{\mu}, x^5, x^6)\equiv \frac{1}{\sqrt{2}}\left(A_6 (x^{\mu}, x^5, x^6) 
+iA_5(x^{\mu}, x^5, x^6)\right),
\end{equation}
which has charge $-1$ under the $SO(2)_{56}$ symmetry. 

In short, in terms of the 4-dimensional ${\cal N}=1$ supersymmetry 
language, the 4-dimensional ${\cal N}=4$ vector 
multiplet consists of one vector multiplet $V=$
$(A_{\mu}, \chi^1)$ and three chiral multiplets 
$\Sigma_1 =(\sigma^1,\chi^2)$, $\Sigma_2=(\sigma^2,\chi^3)$ and  
$\Sigma_3=(\sigma^3,\chi^4)$ in the adjoint representation.

\subsection{ 6-Dimensional ${\cal N} = (1, 1)$ Supersymmetry Breaking
on $T^2$ Orbifolds}

Because we can not keep the 4-dimensional ${\cal N} = 4$ supersymmetry
at low energy from 6-dimensional ${\cal N} = (1, 1)$ supersymmetry,
 we have to break it down to the 4-dimensional ${\cal N} = 1$
 supersymmetry by orbifold projections from the phenomenological point of view.

Before we discuss the 6-dimensional ${\cal N} = (1, 1)$ supersymmetry reaking
on $T^2$ orbifolds, we briefly review the symmetry breaking
on $T^2$ orbifolds. Suppose $\Gamma$ is a gauge (local) or global 
symmetry of the Lagrangian and consider
$Z_n$ $(n=3, 4, 6)$ discrete symmetry on $T^2$, we have 
\begin{eqnarray}
 z \sim \omega  z~,~
{\cal L} (x^{\mu}, \omega z, \omega^{-1} {\bar z})
={\cal L} (x^{\mu}, z, {\bar z}) ~,~\,
\end{eqnarray} 
where $\omega= e^{{\rm i}2\pi/n}$.
So, for a generic bulk multiplet $\Phi$
which fills a representation of the bulk symmetry group $\Gamma$, we have
\begin{eqnarray}
\Phi (x^{\mu}, \omega z, \omega^{-1} {\bar z}) = \eta_{\Phi}
(R_{\omega}^{\Gamma})^{l_\Phi} \Phi (x^{\mu}, z, {\bar z}) 
((R_{\omega}^{\Gamma})^{-1})^{m_\Phi}
~,~\,
\end{eqnarray} 
where $\eta_{\Phi}$, which is a possible phase for the bulk multiplet $\Phi$
from Lagrangian, is an element of $Z_n$.
And $\l_{\Phi}$ and $m_{\Phi}$ are respectively the numbers of
the fundamental index and anti-fundamental index for the bulk multiplet $\Phi$
under the bulk symmetry group $\Gamma$.
 For example, if $\Gamma$ is a 
$SU(N)$ group,
for fundamental representation, $\l_{\Phi}=1$, $m_{\Phi}=0$,
and for adjoint representation, $\l_{\Phi}=1$, $m_{\Phi}=1$. Moreover, 
$R_{\omega}^{\Gamma}$ is the representation of the $Z_n$ symmetry element
$z \rightarrow \omega z$, which
 is introduced to break the bulk symmetry group $\Gamma$.
Also, $R_{\omega}^{\Gamma}$
 is an element in $\Gamma$, and satisfies $(R_{\omega}^{\Gamma})^n = 1$
due to $\omega^n=1$.
Because we disscuss the
supersymmetry breaking and $R$ symmetry breaking in this subsection, 
we choose $\eta_{\Phi}=1$ in the following discussions.

Let us discuss the supersymmetry and $R$ symmetry breakings.
First, we consider the simple case where $R_{\omega}^{SU(2)_{{\bf 4}_+}}$
and $R_{\omega}^{SU(2)_{{\bf 4}_-}}$ are trivial
\begin{eqnarray}
R_{\omega}^{SU(2)_{{\bf 4}_+}} ={\rm diag}[+1, +1] ~,~
R_{\omega}^{SU(2)_{{\bf 4}_-}} ={\rm diag}[+1, +1] ~.~
\end{eqnarray}
Under $Z_n$ symmetry
\begin{eqnarray}
\gamma_n~:~ z \rightarrow \omega  z~,~
\end{eqnarray}
{\it i.e.},  $SO(2)_{56}$ rotation with angle $2\pi/n$,
we obtain that 
\begin{eqnarray}
A_{\mu} (x^{\mu}, \omega z, \omega^{-1} {\bar z}) ~=~ A_{\mu} (x^{\mu}, z, {\bar z})~,~
\sigma^1 (x^{\mu}, \omega z, \omega^{-1} {\bar z}) ~=~ \omega^{-1} 
\sigma^1 (x^{\mu}, z, {\bar z})~,~\,
\end{eqnarray}
\begin{eqnarray}
\sigma^2 (x^{\mu}, \omega z, \omega^{-1} {\bar z}) ~=~  \sigma^2 (x^{\mu}, z, {\bar z})~,~
\sigma^3 (x^{\mu}, \omega z, \omega^{-1} {\bar z}) ~=~  \sigma^3 (x^{\mu}, z, {\bar z})~,~\,
\end{eqnarray} 
\begin{eqnarray}
{\cal Q}_{\alpha, 1}^{(4)} (x^{\mu}, \omega z, \omega^{-1} {\bar z}) ~=~ \omega^{1/2} 
{\cal Q}_{\alpha, 1}^{(4)} (x^{\mu}, z, {\bar z})~,~\,
\label{scharge12}
\end{eqnarray}
\begin{eqnarray}
{\cal Q}_{\alpha, 2}^{(4)} (x^{\mu}, \omega z, \omega^{-1} {\bar z}) ~=~ \omega^{1/2} 
{\cal Q}_{\alpha, 2}^{(4)} (x^{\mu}, z, {\bar z})~,~\,
\end{eqnarray}
\begin{eqnarray}
{\cal Q}_{\alpha, 3}^{(4)} (x^{\mu}, \omega z, \omega^{-1} {\bar z}) ~=~ \omega^{-1/2} 
{\cal Q}_{\alpha, 3}^{(4)} (x^{\mu}, z, {\bar z})~,~\,
\end{eqnarray}
\begin{eqnarray}
{\cal Q}_{\alpha, 4}^{(4)} (x^{\mu}, \omega z, \omega^{-1} {\bar z}) ~=~ \omega^{-1/2} 
{\cal Q}_{\alpha, 4}^{(4)} (x^{\mu}, z, {\bar z})~,~\,
\label{scharge34}
\end{eqnarray}
\begin{eqnarray}
\chi^1 (x^{\mu}, \omega z, \omega^{-1} {\bar z})  ~=~ \omega^{-1/2} 
\chi^1 (x^{\mu}, z, {\bar z}) ~,~
\chi^2 (x^{\mu}, \omega z, \omega^{-1} {\bar z})  ~=~ \omega^{-1/2} 
\chi^2 (x^{\mu}, z, {\bar z}) ~,~\,
\label{gaugino12}
\end{eqnarray}
\begin{eqnarray}
\chi^3 (x^{\mu}, \omega z, \omega^{-1} {\bar z})  ~=~ \omega^{1/2} 
\chi^3 (x^{\mu}, z, {\bar z}) ~,~
\chi^4 (x^{\mu}, \omega z, \omega^{-1} {\bar z})  ~=~ \omega^{1/2} 
\chi^4 (x^{\mu}, z, {\bar z}) ~.~\,
\label{gaugino34}
\end{eqnarray}
From Eqs. (\ref{scharge12}-\ref{gaugino34}), we obtain that
the 4-dimensional ${\cal N} = 4$ supersymmetry is completely broken. 

In order to preserve the 4-dimensional ${\cal N} = 1$ supersymmetry and
make sure that all the bulk fields have KK modes, we choose
\begin{eqnarray}
R_{\omega}^{SU(2)_{{\bf 4}_+}} ={\rm diag}[\omega^{-1/2}, \omega^{+1/2}] ~,~
R_{\omega}^{SU(2)_{{\bf 4}_-}} ={\rm diag}[ \omega^{3/2+k}, \omega^{-3/2-k}] ~,~
\label{Romega}
\end{eqnarray}
where $k$ is an integer, and $0 \leq k < n$.
Then, we have
\begin{eqnarray}
V (x^{\mu}, \omega z, \omega^{-1} {\bar z}) ~=~ V (x^{\mu}, z, {\bar z})~,~
\Sigma_1 (x^{\mu}, \omega z, \omega^{-1} {\bar z}) ~=~ \omega^{-1}
 \Sigma_1 (x^{\mu}, z, {\bar z})~,~\,
\label{gaugetrans1}
\end{eqnarray}
\begin{eqnarray}
\Sigma_2 (x^{\mu}, \omega z, \omega^{-1} {\bar z}) ~=~ \omega^{-1-k} 
\Sigma_2 (x^{\mu}, z, {\bar z})~,~
\Sigma_3 (x^{\mu}, \omega z, \omega^{-1} {\bar z}) ~=~ \omega^{2+k} 
\Sigma_3 (x^{\mu}, z, {\bar z})~.~
\label{gaugetrans2}
\end{eqnarray}
From Eqs. (\ref{gaugetrans1}) and (\ref{gaugetrans2}), we obtain that
for the zero modes, the 4-dimensional ${\cal N} = 4$ supersymmetry
is broken down to the 4-dimensional ${\cal N} = 1$ supersymmetry
 if and only if $1+k \not=0$
mod $n$ and $2+k \not=0$ mod $n$, because in this case
 only the gauge multiplet $V$ has zero modes.

From Eq. (\ref{Romega}),
we obtain that for $k=0$ and $Z_3$ discrete
symmetry ($T^2/Z_3$ orbifold), 
 the $R$ symmetry is $SO(2)_{56}\times U(1)_{{\bf 4}_+}
\times SU(2)_{{\bf 4}_-}$ because $R_{\omega}^{SU(2)_{{\bf 4}_+}}$ breaks
the $ SU(2)_{{\bf 4}_+}$ R symmetry
 down to $U(1)_{{\bf 4}_+}$, and
$R_{\omega}^{SU(2)_{{\bf 4}_-}}$ does not
break the $ SU(2)_{{\bf 4}_-}$ $R$ symmetry. And for the rest cases, the  
$R$ symmetry is $SO(2)_{56}\times U(1)_{{\bf 4}_+} \times  U(1)_{{\bf 4}_-}$
because $R_{\omega}^{SU(2)_{{\bf 4}_+}}$ and $R_{\omega}^{SU(2)_{{\bf 4}_-}}$
breaks the $ SU(2)_{{\bf 4}_+} \times SU(2)_{{\bf 4}_-}$ $R$ symmetry
down to $U(1)_{{\bf 4}_+} \times  U(1)_{{\bf 4}_-}$.

In addition, to preserve the 4-dimensional ${\cal N} = 1$ supersymmetry,
we find that for the $T^2/Z_3$ orbifold,
the only possibility is $k=0$. And in this case, the
$R$ symmetry is $SO(2)_{56}\times U(1)_{{\bf 4}_+}
\times SU(2)_{{\bf 4}_-}$. Moreover,
for the $T^2/Z_4$ orbifold, $k=0$ is equivalent to
$k=1$, {\it i.e.}, there is also only one possibility. 
 For the $T^2/Z_6$ orbifold, the inequivalent cases
are $k=0$ and $k=1$. The $R$ symmetry for  $T^2/Z_4$
and $T^2/Z_6$ orbifolds is $SO(2)_{56}\times U(1)_{{\bf 4}_+} 
\times  U(1)_{{\bf 4}_-}$.

The maximal $R$ symmetry for
the 4-dimensional ${\cal N} = 4$ supersymmetry is $SU(4)_R$. 
Embedding the $SO(2)_{56}\times SU(2)_{{\bf 4}_+} \times SU(2)_{{\bf 4}_-}$
$R$ symmetry into the $SU(4)_R$ gives us an elegant way to describe the
supersymmetry and $R$ symmetry breakings, so, we study it in detail.

The generator of $SO(2)_{56}$ in $SU(4)_R$ is
\begin{eqnarray}
T_{SO(2)_{56}} ={\rm diag}\left[+{1\over 2}, +{1\over 2}, -{1\over 2}, 
-{1\over 2}\right]~,~\,
\label{GSO2R}
\end{eqnarray}
and the generators of $SU(2)_{{\bf 4}_+} \times SU(2)_{{\bf 4}_-}$ are embedded
into $SU(4)_R$ as
\begin{eqnarray}
\left( \begin{array}{cc}
 T_{SU(2)_{\bf 4_+}}& 0 \\ 0& T_{SU(2)_{\bf 4_-}}
       \end{array}\right) \subset 
\left(T_{SU(4)_{ R}}\right)~.~\,
\label{GSU2+-}
\end{eqnarray}

We define
\begin{eqnarray}
\left({\cal Q}_\alpha^{(4)}\right)^T \equiv \left(
{\cal Q}_{\alpha, 1}^{(4)}, ~{\cal Q}_{\alpha, 2}^{(4)},
~{\cal Q}_{\alpha, 3}^{(4)}, ~{\cal Q}_{\alpha, 4}^{(4)} \right)~,~\,
\end{eqnarray}
\begin{eqnarray}
\left(\chi_{\alpha}\right)^T \equiv \left( \chi_{\alpha}^1,
~\chi_{\alpha}^2, ~\chi_{\alpha}^3, ~\chi_{\alpha}^4\right)~,~\,
\end{eqnarray}
\begin{eqnarray}
 \sigma^{ab} \equiv \left(\begin{array}{cccc}
		0 & \sigma^1 & \sigma^2 & \sigma^3 \\
                -\sigma^1 & 0 & \sigma^{3*} & -\sigma^{2*} \\
                -\sigma^2 & -\sigma^{3*} & 0 & \sigma^{1*} \\
                -\sigma^3 & \sigma^{2*} & -\sigma^{1*} & 0 \\
		     \end{array}\right)~.~\,
\end{eqnarray}
And then,
${\cal Q}_\alpha^{(4)}$ transforms as the fundamental representation ${\bf 4}$ under
$SU(4)_R$, $\chi_{\alpha}$ transforms as the anti-fundamental representation 
${\bf {\bar 4}}$ under $SU(4)_R$, and $\sigma^{ab}$ transforms as the anti-symmetric  
tensor ${\bf {\bar 4}\wedge {\bar 4}}$ of two anti-fundamental representations
under $SU(4)_R$.

Similar to the weakly-coupled heterotic string compactification, in order
to preserve the 4-dimensional ${\cal N} = 1$ supersymmetry, $R_{\omega}^{SU(4)_R}$
should take the following form
\begin{eqnarray}
R_{\omega}^{SU(4)_R} =\left(\begin{array}{cc}
1 & 0 \\
0 & R_{\omega}^{SU(3)_R}\\
		     \end{array}\right)~.~\,
\end{eqnarray}
The supercharges $\left({\cal Q}_{\alpha, 2}^{(4)},
~{\cal Q}_{\alpha, 3}^{(4)}, ~{\cal Q}_{\alpha, 4}^{(4)}\right)^T$ form a fundamental 
representation under the $SU(3)_R$,
$\left(\chi_{\alpha}^2, ~\chi_{\alpha}^3, ~\chi_{\alpha}^4\right)^T$ and
$\left(\sigma^1, ~\sigma^2, ~\sigma^3\right)^T$ form the anti-fundamental 
representation under the $SU(3)_R$. 

With Eqs. (\ref{Romega}), (\ref{GSO2R}) and (\ref{GSU2+-}),
 we obtain 
\begin{eqnarray}
R_{\omega}^{SU(4)_R} &=& {\rm diag}[\omega^{+1/2}, \omega^{+1/2}, 
\omega^{-1/2}, \omega^{-1/2}]\times {\rm diag}[\omega^{-1/2}, \omega^{+1/2}, 1, 1]
\nonumber\\ &&
\times {\rm diag}[1, 1, \omega^{3/2+k}, \omega^{-3/2-k}]~.~\,
\end{eqnarray}
Thus, we have 
\begin{eqnarray}
R_{\omega}^{SU(4)_R} &=& {\rm diag}[+1, \omega, \omega^{1+k}, \omega^{-2-k}]~.~\,
\end{eqnarray}
Similar to above discussions, 
for the zero modes, the 4-dimensional ${\cal N} = 4$ supersymmetry
is broken down to the 4-dimensional ${\cal N} = 1$ supersymmetry
if and only if $1+k \not=0$
mod $n$ and $2+k \not=0$ mod $n$.
And the transformations of vector multiplet under $Z_n$ are
given by Eqs. (\ref{gaugetrans1}) and (\ref{gaugetrans2}).

Suppose $G$ is a Lie group and $H$ is a subgoup of $G$, for our convention,
we denote the commutant of $H$ in $G$ as $G/H$, {\it i.e.},
\begin{equation}
G/H\equiv \{g \in G|gh=hg, ~{\rm for ~any} ~ h \in H\}~.~\,
\end{equation}

With this convention, the unbroken $R$ symmetry is given by
\begin{eqnarray}
(SU(4)_R/R_{\omega}^{SU(4)_R}) ~\cap ~ 
(SO(2)_{56}\times SU(2)_{{\bf 4}_+} \times SU(2)_{{\bf 4}_-})~.~\,
\end{eqnarray}
Similar to above results, to preserve the 4-dimensional ${\cal N} = 1$ supersymmetry,
we obtain that for $T^2/Z_3$ orbifold, the only possibility is $k=0$ and the
$R$ symmetry is $SO(2)_{56}\times U(1)_{{\bf 4}_+}
\times SU(2)_{{\bf 4}_-}$. And for $T^2/Z_4$ orbifold, the only possibility is
$k=0$, too. For  
the $T^2/Z_6$ orbifold, there are two possibilities: $k=0$ and $k=1$.  
The $R$ symmetry for  $T^2/Z_4$
and $T^2/Z_6$ orbifolds is $SO(2)_{56}\times U(1)_{{\bf 4}_+} 
\times  U(1)_{{\bf 4}_-}$.

To be concrete, we define the generators for $U(1)_{{\bf 4}_+}$
 and $U(1)_{{\bf 4}_-}$ in the $SU(4)_R$
\begin{eqnarray}
T_{U(1)_{{\bf 4}_+}} \equiv {\rm diag}\left[+{1\over 2}, -{1\over 2}, 0, 0\right]~,~\,
\end{eqnarray}
\begin{eqnarray}
T_{U(1)_{{\bf 4}_-}} \equiv {\rm diag}\left[0, 0, +{1\over 2}, -{1\over 2}\right]~.~\,
\end{eqnarray}

To be complete, for the bulk gauge group $G$, we write down the  bulk action 
in the Wess-Zumino gauge and 4-dimensional ${\cal N}=1$ supersymmetry
language~\cite{NMASWS, NAHGW}
\begin{eqnarray}
  S &=& \int d^6 x \Biggl\{
  {\rm Tr} \Biggl[ \int d^2\theta \left( \frac{1}{4 k g^2} 
  {\cal W}^\alpha {\cal W}_\alpha + \frac{1}{k g^2} 
  \left( \Sigma_3 \partial \Sigma_2   - \frac{1}{\sqrt{2}} \Sigma_1 
  [\Sigma_2, \Sigma_3] \right) \right) + {\rm h.c.} \Biggr] 
\nonumber\\
  && + \int d^4\theta \frac{1}{k g^2} {\rm Tr} \Biggl[ 
  (\sqrt{2} \partial^\dagger + \Sigma_1^\dagger) e^{-V} 
  (-\sqrt{2} \partial + \Sigma_1) e^{V}\Biggr]
\nonumber\\
&&+ \int d^4\theta \frac{1}{k g^2} {\rm Tr} \Biggl[
   \Sigma_2^\dagger e^{-V} \Sigma_2  e^{V}
  + {\Sigma_3}^\dagger e^{-V} \Sigma_3 e^{V} 
\Biggr] \Biggr\}.
\label{eq:t2z6action}
\end{eqnarray}
From above action, we obtain  
the transformations of vector multiplet 
\begin{eqnarray}
  V(x^{\mu}, ~\omega z, ~\omega^{-1} {\bar z}) &=& (R_{\omega}^G)^{l_V}
 V(x^{\mu}, ~z, ~{\bar z}) ((R_{\omega}^G)^{-1})^{m_V}~,~\,
\label{Vtrans}
\end{eqnarray}
\begin{eqnarray}
  \Sigma_1(x^{\mu}, ~\omega z, ~\omega^{-1} {\bar z}) &=& 
\omega^{-1} (R_{\omega}^G)^{l_{\Sigma_1}} 
\Sigma_1(x^{\mu}, ~z, ~{\bar z}) ((R_{\omega}^G)^{-1})^{m_{\Sigma_1}}~,~\,
\label{S1trans}
\end{eqnarray}
\begin{eqnarray}
   \Sigma_2(x^{\mu}, ~\omega z, ~\omega^{-1} {\bar z}) &=& 
\omega^{-1-k} (R_{\omega}^G)^{l_{\Sigma_2}} 
\Sigma_2(x^{\mu}, ~z, ~{\bar z})  ((R_{\omega}^G)^{-1})^{m_{\Sigma_2}}~,~\,
\label{S2trans}
\end{eqnarray}
\begin{eqnarray}
 \Sigma_3(x^{\mu}, ~\omega z, ~\omega^{-1} {\bar z})  &=& 
\omega^{2+k} (R_{\omega}^G)^{l_{\Sigma_3}}  
\Sigma_3(x^{\mu}, ~z, ~{\bar z}) ((R_{\omega}^G)^{-1})^{m_{\Sigma_3}}~,~\,
\label{S3trans}
\end{eqnarray}
where we introduce the non-trivial
$R_{\omega}^G$ to break the bulk gauge group $G$.
When we consider the gauge symmetry breaking by discrete symmetry
on the extra space manifold, we can
decompose the gauge fields of $G$ under its maximal subgroup, and then,
discuss the gauge symmetry breaking, for example, the $E_6$ breaking and $E_8$
breaking in Refs.~\cite{HJLL} 
and~\cite{JUNII}, respectively. Some of the gauge fields of $G$ are not in the
adjoint representations under its maximal subgroup due to decomposition,
so, we write down the explicit $l_{\Phi}$ and $m_{\Phi}$ for the general orbifold
gauge symmetry breaking.

Moreover, under the $SO(2)_{56}$ $R$ symmetry, the $\theta$ (the Grassmann
coordinate for the unbroken 4-dimensional ${\cal N}=1$ supersymmetry), 
$z$ and the bulk vector multiplet transform as
\begin{eqnarray}
z \longrightarrow e^{{\rm i} \beta } z
~,~ \theta \longrightarrow e^{-{\rm i} \beta/2} \theta ~,~
V \longrightarrow  V ~,~\,
\end{eqnarray}
\begin{eqnarray}
\Sigma_1 \longrightarrow e^{-{\rm i} \beta } \Sigma_1 ~,~
\Sigma_2 \longrightarrow  \Sigma_2 ~,~
\Sigma_3 \longrightarrow  \Sigma_3 ~.~\,
\end{eqnarray}

Under the $U(1)_{{\bf 4}_+}$ $R$ symmetry, the Grassmann
coordinate $\theta$, 
$z$ and the bulk vector multiplet transform as
\begin{eqnarray}
z \longrightarrow z
~,~ \theta \longrightarrow e^{-{\rm i} \beta/2} \theta ~,~
V \longrightarrow  V ~,~\,
\end{eqnarray}
\begin{eqnarray}
\Sigma_1 \longrightarrow  \Sigma_1 ~,~
\Sigma_2 \longrightarrow e^{-{\rm i} \beta/2 } \Sigma_2 ~,~
\Sigma_3 \longrightarrow e^{-{\rm i} \beta/2 } \Sigma_3 ~.~\,
\end{eqnarray}

For the $T^2/Z_3$ orbifold,  $(\Sigma_2, \Sigma_3)^T$
form an anti-fundamental representation under $SU(2)_{{\bf 4}_-}$.
And for the $T^2/Z_4$ and $T^2/Z_6$ orbifolds, $\Sigma_2$ and $\Sigma_3$
have the $U(1)_{{\bf 4}_-}$ charges $-1/2$ and $+1/2$, respectively,
while $\theta$, $z$, $V$ and $\Sigma_1$ are neutral.

\section{${\cal N} = (1, 1)$ Supersymmetric $E_6$ Models on $M^4\times T^2/Z_3$}

In this Section, we discuss the
${\cal N} = (1, 1)$ supersymmetric $E_6$ models
 on the space-time $M^4\times T^2/Z_3$ where the gauge symmetry
$E_6$ is broken down to $SO(10)\times U(1)_X$ by orbifold projection. 
And there are three $SO(10)$ spinor representation ${\bf 16}$s from the
zero modes of bulk vector multiplet, which can be considered
as three families of fermions. We will discuss two models in detail, and
show that the fermion masses and mixings can be explained by the flavour
symmetry $SU(2)\times U(1)$ whose origin is $R$ symmetry. 
By the way, in this Section, we consider the $T^2/Z_3$ orbifold, 
so, $\omega=e^{{\rm i} 2\pi/3}$.

The gauge fields of $E_6$ are in the adjoint representation 
 with dimension {\bf 78}. Under the gauge group
$SO(10)\times U(1)_X$, the $E_6$ gauge fields
decompose as~\cite{Group}
\begin{eqnarray}
{\bf 78= (45, 0) \oplus (16, -3) \oplus (\overline{16}, 3) \oplus
(1, 0)}~.~\,
\end{eqnarray} 

The $R_{\omega}^{E_6}$ for $E_6$ gauge group is the product of the 
$R_{\omega}^{SO(10)}$ for $SO(10)$ and $R_{\omega}^{U(1)_X}$ for
$U(1)_X$
\begin{eqnarray}
R_{\omega}^{E_6} = R_{\omega}^{SO(10)} \otimes R_{\omega}^{U(1)_X}~.~\,
\label{RE6}
\end{eqnarray}
In order to break the $E_6$ gauge symmetry, we choose 
\begin{eqnarray}
R_{\omega}^{SO(10)} &=& 
{\rm diag}[+1, +1, +1, +1, +1, +1, +1, +1, +1, +1]~,~\,
\label{RSO10}
\end{eqnarray}
\begin{eqnarray}
R_{\omega}^{U(1)_X}  &=& {\rm exp}\left(-{\rm i} 2 \pi Q/9\right)~,~\,
\label{RU(1)}
\end{eqnarray}
where $Q$ is the $U(1)_X$ charge of the field. And the sufficient condition
for $R_{\omega}^{U(1)_X}$ to satisfy is that $(R_{\omega}^{U(1)_X})^3$ is
equal to $1$ for all the bulk fields which are charged under $U(1)_X$.

Using Eqs. (\ref{Vtrans}-\ref{S3trans}) with $k=0$ and Eqs. 
(\ref{RE6}-\ref{RU(1)}), we obtain the transformations of vector multiplet 
\begin{eqnarray}
  V_{{\bf (45, 0)}} (x^{\mu}, ~\omega z, ~\omega^{-1} {\bar z}) ~=~ 
 V_{{\bf (45, 0)}} (x^{\mu}, ~z, ~{\bar z})  ~,~\,
\label{VV1}
\end{eqnarray}
\begin{eqnarray}
V_{{\bf (1, 0)}} (x^{\mu}, ~\omega z, ~\omega^{-1} {\bar z}) ~=~ 
 V_{{\bf (1, 0)}} (x^{\mu}, ~z, ~{\bar z})~,~\,
\label{VV2}
\end{eqnarray}
\begin{eqnarray}
  V_{{\bf (16, -3)}} (x^{\mu}, ~\omega z, ~\omega^{-1} {\bar z}) ~=~ \omega
 V_{{\bf (16, -3)}} (x^{\mu}, ~z, ~{\bar z}) ~,~\,
\label{VV3}
\end{eqnarray}
\begin{eqnarray} 
V_{{\bf (\overline{16}, 3)}} (x^{\mu}, ~\omega z, ~\omega^{-1} {\bar z}) ~=~ 
\omega^2 V_{{\bf (\overline{16}, 3)}} (x^{\mu}, ~z, ~{\bar z})~,~\,
\label{VV4}
\end{eqnarray}
\begin{eqnarray}
  (\Sigma_i)_{{\bf (45, 0)}} (x^{\mu}, ~\omega z, ~\omega^{-1} {\bar z}) ~=~ \omega^2
(\Sigma_i) _{{\bf (45, 0)}} (x^{\mu}, ~z, ~{\bar z}) ~,~\,  
\label{SS1}
\end{eqnarray}
\begin{eqnarray}
(\Sigma_i)_{{\bf (1, 0)}} (x^{\mu}, ~\omega z, ~\omega^{-1} {\bar z}) ~=~ \omega^2
  (\Sigma_i)_{{\bf (1, 0)}} (x^{\mu}, ~z, ~{\bar z})~,~\,
\label{SS2}
\end{eqnarray}
\begin{eqnarray}
  (\Sigma_i)_{{\bf (16, -3)}} (x^{\mu}, ~\omega z, ~\omega^{-1} {\bar z}) ~=~ 
 (\Sigma_i)_{{\bf (16, -3)}} (x^{\mu}, ~z, ~{\bar z}) ~,~\, 
\label{SS3}
\end{eqnarray}
\begin{eqnarray}
(\Sigma_i)_{{\bf (\overline{16}, 3)}} (x^{\mu}, ~\omega z, ~\omega^{-1} {\bar z}) ~=~ 
\omega (\Sigma_i)_{{\bf (\overline{16}, 3)}} (x^{\mu}, ~z, ~ {\bar z})~,~\,
\label{SS4}
\end{eqnarray}
where $i=1, 2, 3$, and $\omega=e^{{\rm i} 2\pi/3}$.

From the general discussions on gauge symmetry breaking by discrete
symmetry~\cite{JUNII}, we obtain that only the fields
$V_{{\bf (45, 0)}}$, $V_{{\bf (1, 0)}}$, and
 $(\Sigma_i)_{{\bf (16, -3)}}$ have zero modes. So, the
$E_6$ gauge symmetry is broken down to $SO(10)\times U(1)_X$
by orbifold projection for the zero modes. 
And there are three $SO(10)$ spinor representations ${\bf 16}$ from the
zero modes of bulk vector multiplet, which can be considered
as three families of fermions. 
This may explain why there are three families
of fermions in the nature. Moreover,
the $R$ symmetry is $SO(2)_{56}\times U(1)_{{\bf 4}_+}
\times SU(2)_{{\bf 4}_-}$, which can be the origin of
flavour symmetry to generate the fermion masses and mixings. 

\subsection{Model I}
In  model I,  the zero modes $ (\Sigma_i)^{(0)}_{{\bf (16, -3)}}$
 of $ (\Sigma_i)_{{\bf (16, -3)}}$ where
$i=1, 2, 3$ are considered as three families of fermions, and the 
 $SO(2)_{56}\times U(1)_{{\bf 4}_+}\times SU(2)_{{\bf 4}_-}$ $R$ symmetry 
can give us the flavour symmetry to generate the fermion masses and mixings.

To be convenient in discussions, we define 
\begin{eqnarray}
{\bf 16}_1 \equiv - (\Sigma_3)^{(0)}_{{\bf (16, -3)}}
~,~ {\bf 16}_2 \equiv  (\Sigma_2)^{(0)}_{{\bf (16, -3)}}
~,~ {\bf 16}_3 \equiv  (\Sigma_1)^{(0)}_{{\bf (16, -3)}}~,~\, 
\end{eqnarray}
where ${\bf 16}_i$ is the $i-th$ family of the Standard Model fermions.
In this convention, the first two families
$({\bf 16}_1, {\bf 16}_2)^T$ form a fundamental
representation under the $R$ symmetry $SU(2)_{{\bf 4}_-}$.

In order to give the same $U(1)_R$ charges for three families of fermions 
${\bf 16}_i$, we define the $U(1)_R^{I}$ and $U(1)_F^{I}$ global symmetries from  
the $SO(2)_{56} \times U(1)_{{\bf 4}_+}$ $R$ symmetry, whose generators in
$SU(4)_R$ are
\begin{eqnarray}
T_{U(1)_R^{I}} ={\rm diag}\left[+{3\over 2}, -{1\over 2}, -{1\over 2}, 
-{1\over 2}\right]~,~\,
\label{IU1R}
\end{eqnarray}
\begin{eqnarray}
T_{U(1)_F^{I}} ={\rm diag}\left[0, -1, +{1\over 2}, +{1\over 2}\right]~.~\,
\end{eqnarray}
So, under $U(1)_R^{I}$, the Grassmann coordinate $\theta$ has charge
$-3/2$, and three families of fermions (superfields)
${\bf 16}_i$ have the same charge $-1$. Then, the superpotential has
$U(1)_R^{I}$ charge $-3$. In addition, under $U(1)_F^{I}$,
the Grassmann coordinate $\theta$ has charge 0, 
${\bf 16}_3$ has charge $+1$, and ${\bf 16}_1$ and ${\bf 16}_2$ have
charge $-1/2$. 

In Model I, the $R$ symmetry $U(1)_F^{I} \times SU(2)_{{\bf 4}_-}$
can be considered as flavour symmetry.
People have discussed the fermion masses and
mixings extensively in the supersymmetric $SO(10)$ models with $U(2)$ or
 $SU(2)\times U(1)^n$ flavour 
symmetries~\cite{FSU2, SRABY1, SRABY2, MCKTM}. In this
paper, we follow the discussions in Ref.~\cite{SRABY1}, where the flavour symmetry
is $SU(2)\times U(1)$ and the fermion masses and
mixings can be explained.
 
To simplify the convention, we denote ${\bf 16}_1$ and ${\bf 16}_2$
as ${\bf 16}_{a'}$ where the low index $a'$ is the $SU(2)_{{\bf 4}_-}$ index
for fundamental representation. And the upper index $a'$ is the
$SU(2)_{{\bf 4}_-}$ index for anti-fundamental representation.
We also assume that the particles, which are introduced to break the gauge symmetry
and generate the fermion masses and mixings, are on 
the observable 3-brane at the fixed point $z=0$.
Similar to the particle content in Ref.~\cite{SRABY1}, we introduce
a ${\bf 10}$ dimensional Higgs field $H$, a ${\bf 45}$ dimensional Higgs
 field ${\bf 45}_H$ which has a VEV along the $B-L$ direction, and
one pair of ${\bf 16}$ and $\overline{\bf 16}$ Higgs fields
(${\psi}_H$, $\overline{\psi}_H$)
which are assumed to have the VEVs along the ``right-handed'' neutrino direction
and break the $SO(10)$ gauge symmetry down to $SU(5)$. $\overline{\psi}_H$
can generate the right-handed neutrino Majorana masses. Moreover, we introduce
three pairs of heavy ${\bf 16}$s and $\overline{\bf 16}$s of $SO(10)$:
($\psi$, $\overline{\psi}$), ($\psi^{a'}$, $\overline{\psi}_{a'}$),
($\psi_{a'}$, $\overline{\psi}^{a'}$). These heavy ${\bf 16}$s and $\overline{{\bf 16}}$s
do not have VEVs, and we can obtain the non-renormalizable Yukawa matrices
after we integrate them out. We also add three
$SO(10)$ singlets $N_{a'}$ and $N_3$.

In order to break the flavour symmetry $U(1)_F^{I} \times SU(2)_{{\bf 4}_-}$,
we introduce an anti-fundamental representation $\phi^{a'}$ (doublet), a
symmetric representation with two anti-fundamental indices $S^{a' b'}$ (triplet),
and an anti-symmetric representation with two anti-fundamental indices 
$A^{a' b'}$ (singlet)
 where the representations are under $SU(2)_{{\bf 4}_-}$ symmetry.
In addition, $\phi^{a'}$, $S^{a' b'}$, and $A^{a' b'}$ are $SO(10)$
singlets and charged under $U(1)_F^{I}$.

The quantum numbers for above particles in Model I under $U(1)_X$ gauge
symmetry and $U(1)_R^{I}\times U(1)_F^{I} $
$R$ symmetry are given in Table \ref{spectrumI}
where $Q_X$, $Q_R$ and $Q_F$ are free parameters.\\

I. Low Energy $R$ Symmetry.\\

As we know, under the
traditional $U(1)_R$ symmetry defined in
the low energy supersymmetry phenomenology, the Grassmann coordinate $\theta$
and the superfields ${\bf 16}_i$ have charge $+1$, while the Higgs fields ($H$
and ${\bf 45}_H$) have
charge zero. So, the
$U(1)_R^I$ symmetry defined in Eq. (\ref{IU1R})
 is not the traditional $U(1)_R$ symmetry, and it is broken by
the VEVs of $H$, ${\psi}_H$ and $\overline{\psi}_H$ (The VEVs of 
$\phi^{a'}$, $S^{a' b'}$, and $A^{a' b'}$ will not break the $U(1)_R^I$ symmetry
if $Q_R=0$.). Moreover, there is no unbroken 
$Z_2$ symmetry under
which the ordinary particles are even while their superpartners
are odd. Thus, there are no cold dark matter candidates (for example neutralino)
from the superpartners, although the axion can still be the cold dark matter
candidate. However, this $U(1)_R^I$ symmetry can still forbid the dimension-5 proton
decay operators ${\bf 16}_i {\bf 16}_j {\bf 16}_k {\bf 16}_l$. 
And we would like to point out
that this $U(1)_R^I$ symmetry may be the unbroken $U(1)$ R symmetry 
in the compactification of the weakly coupled
heterotic string theory on Calabi-Yau manifolds. So, it is interesting to
study its low energy phenomenology. In the next subsection, we will discuss
a model where the first two families come from the zero modes of bulk vector
multiplet and the $U(1)$ $R$ symmetry is the traditional $U(1)_R$ $R$ symmetry
at low energy supersymmetry phenomenology.\\

II. Anomaly Cancellations.\\

In our model, there is no $[SO(10)]^3$ anomaly. But, there exist the mixed
anomalies $[SO(10)]^2 U(1)_X$, $[U(1)_X]^3$, and $[{\rm Gravity}]^2 U(1)_X$,
which are localized on three 3-branes at three fixed points. All three 
anomalies can be cancelled by a generalized Green-Schwarz
 mechanism~\cite{GSM}, and then,
$U(1)_X$ becomes a global symmetry which should be preserved by the superpotential
localized on the 3-branes at the fixed points. However, the Fayet-Iliopoulos 
D-terms may be generated radiatively on three 3-branes at three fixed 
points~\cite{DSEW}. To avoid the supersymmetry breaking from these
Fayet-Iliopoulos D-terms, we require the $U(1)_X$ D-flatnesses on 
three 3-branes at three fixed points.
And to achieve the $U(1)_X$ D-flatnesses,
 we introduce one pair of the $SO(10)$ singlets with $U(1)_X$ charges $\pm1$ on 
each 3-brane, and give them non-zero VEVs.
Therefore, if we assume that
the anomalies are cancelled by a generalized Green-Schwarz mechanism,
we do not need to consider the $U(1)_X$ anomalies and 
its breaking, although we 
require that the superpotential localized on the 3-branes at the fixed points
 preserve the $U(1)_X$ symmetry. 

In addition, we can cancel the anomalies by introducing the vector-like 
particles on the 3-branes at the fixed points. The anomalies due to the bulk
vector multiplets  arise from the zero modes of 
$(\Sigma_i)_{{\bf (16, -3)}}$, {\it i.e.}, three families of fermions
${\bf 16}_{a'}$ and ${\bf 16}_3$. And these anomalies distribute equally 
on three 3-branes at
three fixed points since these fixed points are equivalent from geometry.
Therefore, the anomalies on the 3-brane at each fixed point due to the bulk
vector multiplet are equivalent to the anomalies from a multiplet
with quantum number $({\bf 16, -3})$ under the $SO(10)\times U(1)_X$
gauge symmetry. And we just need to cancel the anomalies on the 3-brane at
each fixed point.

To simplify the discussions, we choose $Q_X=0$. First, 
on the observable 3-brane at $z=0$, 
the anomalies due to the bulk vector multiplet
are cancelled by the Higgs field $\overline{\psi}_H$. Then, we
only need to cancel the anomalies generated by the Higgs field $H$.
So, we introduce one pair of 10-plets $T_{10}$  and $T_{10}'$ whose
 quantum numbers are $({\bf 10, -6})$ and $({\bf 10, 0})$ respectively
under the $SO(10)\times U(1)_X$ gauge symmetry. In order to give the masses
to $T_{10}$  and $T_{10}'$, and keep
the $U(1)_X$  D-flatness and anomaly free on the 3-brane at $z=0$,
we introduce the
$SO(10)$ singlets $s_1$ and $s'_1$ with $U(1)_X$ charges $+6$ and $-6$,
respectively. In order to forbid the mixings between the Higgs 
$H$ and $T_{10}$ or between $H$ and $T_{10}'$, we assign the following
$U(1)_R^{I}\times U(1)_F^{I}$ charges for these vector-like particles: the
$U(1)_R^{I}$ charges for $T_{10}$, $T_{10}'$, $s_1$ and $s'_1$
are 0, 0, $-3$ and 0, respectively, and the $U(1)_F^{I}$ charges
for $T_{10}$, $T_{10}'$, $s_1$ and $s'_1$ are $-1$, $-1$, 2 and
$-1$, respectively. One can check that the 
anomalies $[SO(10)]^2 U(1)_X$, $[U(1)_X]^3$, and $[{\rm Gravity}]^2 U(1)_X$
are cancelled on the 3-brane at $z=0$. And the masses of 
$T_{10}$  and $T_{10}'$ can be generated by the superpotential
$s_1 T_{10} T_{10}'$.

Second, on the 3-brane at $z=\pi R e^{{\rm i}\pi/6}/{\sqrt 3}$,
we need to cancel the anomalies generated by a multiplet
with quantum number $({\bf 16, -3})$ under the $SO(10)\times U(1)_X$
gauge symmetry. So, we introduce a pair of the 
$\Upsilon_2$ and $\Upsilon_2'$  whose
 quantum numbers are $({\bf 16, 0})$ and $({\bf \overline{16}, +3})$ respectively
under the $SO(10)\times U(1)_X$ gauge symmetry. And in order to give the masses
to $\Upsilon_2$  and $\Upsilon_2'$, and keep
the $U(1)_X$ D-flatness and anomaly free on the 3-brane 
at $z=\pi R e^{{\rm i}\pi/6}/{\sqrt 3}$, we introduce the
$SO(10)$ singlets $s_2$ and $s'_2$ with $U(1)_X$ charges $-3$ and $+3$,
respectively.
 To forbid the mixings between the SM fermions 
${\bf 16}_i$ and $\Upsilon_2$ and the mass terms between
 $ {\bf 16}_i$ and $\Upsilon_2'$, we assign the following
$U(1)_R^{I}\times U(1)_F^{I}$ charges for the vector-like particles: the
$U(1)_R^{I}$ charges for $\Upsilon_2$, $\Upsilon_2'$, $s_2$ and $s'_2$
are 0, 0, $-3$ and 0, respectively, and the $U(1)_F^{I}$ charges
for $\Upsilon_2$, $\Upsilon_2'$, $s_2$ and $s'_2$ are $-1/2$, $-1/2$, 1 and
$-1$, respectively. In short, the 
anomalies $[SO(10)]^2 U(1)_X$, $[U(1)_X]^3$, and $[{\rm Gravity}]^2 U(1)_X$
are cancelled on the 3-brane at $z=\pi R e^{{\rm i}\pi/6}/{\sqrt 3}$. 
And the masses of 
 $\Upsilon_2$  and $\Upsilon_2'$ can be generated by the superpotential
$s_2 \Upsilon_2 \Upsilon_2'$.

Third,  the anomaly cancellations on the 3-brane at 
$z= 2 \pi R e^{{\rm i}\pi/6}/{\sqrt 3}$ is similar to those
on the 3-brane at $z=\pi R e^{{\rm i}\pi/6}/{\sqrt 3}$. We introduce
the particles $\Upsilon_3$, $\Upsilon_3'$, $s_3$ and $s'_3$, whose
quantum numbers are the same as those for
$\Upsilon_2$, $\Upsilon_2'$, $s_2$ and $s'_2$, respectively.

Furthermore, similar to the 5-dimensional orbifold theories~\cite{anomaly}
and string theory~\cite{top}, it seems to us 
that there exist the anomaly inflows among the 3-branes at the fixed points.
If this is true, like the 5-dimensional orbifold theories,
the 4-dimensional anomaly cancellations are sufficient
 to ensure the consistency of higher dimensional orbifold theories~\cite{anomaly}.
In other words, we only need to consider the total anomalies due to
the bulk zero modes and the particles localized on the 3-branes at
the fixed points. For example, we can put 
$\Upsilon_2$, $\Upsilon_2'$, $s_2$, $s'_2$, 
$\Upsilon_3$ and $\Upsilon_3'$  on the 3-brane at $z=\pi R e^{{\rm i}\pi/6}/{\sqrt 3}$.
The anomalies are cancelled due to anomaly inflows, 
and $\Upsilon_3$ and $\Upsilon_3'$
can obtain the masses via the superpotential $s_2 \Upsilon_3 \Upsilon_3'$.
The explicit proof of the anomaly inflows among the 3-branes at the fixed points
on $T^2$ orbifolds is an interesting open question.\\

III. Fermion Masses and Mixings.\\

The general VEVs for $\phi^{a'}$, $S^{a' b'}$, and $A^{a' b'}$, which break
the $U(1)_F^{I} \times SU(2)_{{\bf 4}_-}$ flavour symmetry, are
\begin{eqnarray}
\langle \phi^2 \rangle  
= r \, \langle S^{2 2} \rangle ~,~\langle A^{1 2} \rangle  \neq 0
~,~\, 
\end{eqnarray}
\begin{eqnarray}
 \langle S^{2 2} \rangle  \neq 0~, &  
\langle S^{1 1} \rangle = \kappa_1 \, \langle S^{2 2} \rangle ~,~ 
\langle S^{1 2} \rangle = \kappa_2 \, \langle S^{2 2} \rangle~,~ \, 
\end{eqnarray}
where the constants $r$, $\kappa_1$ and $\kappa_2$ are arbitrary.  

The superpotential is 
\begin{eqnarray}
 W  &=&  16_3 \; H\; 16_3 \;\; +\; \;  16_{a'} \; H  \; \psi^{a'} 
+ \;\; \overline{\psi} \; (M' \; \psi \;\; + \;\; 45_H \; 16_3)
\nonumber\\ &&
 + \; \overline{\psi}_{a'}
 \; (M \; \psi^{a'} \; +\; \phi^{a'} \; \psi \; +\; S^{a'\; b'} \;
 \psi_{b'} \; + \;  A^{a'\; b'} \; 16_{b'}) 
\nonumber\\ &&
+ \;\; \overline{\psi}^{a'} \; (M'' \; \psi_{a'} \;\; + \;\; 45_H \; 16_{a'}) 
+ \overline{\psi}_H \; (N_{a'} \; \psi^{a'} \;\; + \;\; N_3 \; 16_3) 
\nonumber\\ &&
 + {1 \over 2} \ N_{a'} \; N_{b'} \; 
S^{a' \; b'} \;\; + \;\; N_{a'} \; N_3 \; \phi^{a'} ~,~\,
\label{superpotential}
\end{eqnarray}
where 
\begin{equation}
M = M_0 \left(1\;\;+ \;\;  
\alpha_{X'} \; T_{X'} \;\; +\;\; \beta_Y \; 
 T_Y \right)~,~
\label{eq:M0}
\end{equation}
\begin{eqnarray}
\alpha_{X'} \propto {{\langle {\bf 45}_{X'} \rangle} \over M_0}~,~
\beta_{Y} \propto {{\langle {\bf 45}_{Y} \rangle} \over M_0}~,~\,
\end{eqnarray}
where $T_{X'}$ and $T_Y$ are the generators for $U(1)_{X'}$ and 
$U(1)_{Y}$, respectively where the
$U(1)_{X'}$ is the $U(1)$ in $SO(10)$ which commutes with $SU(5)$,
and the $U(1)_{Y}$ is the standard weak hypercharge interaction.
The Higgs fields $ {\bf 45}_{X'}$ and $ {\bf 45}_Y $ have VEVs
along the $U(1)_{X'}$ and $U(1)_{Y}$ directions, respectively. 
And $\alpha_{X'}$ and $\beta_Y$ are arbitrary constants which
are used to fit the data. Futhermore, each term in superpotential
has an order $1$ (${\cal O} (1)$) coupling constant which
is omitted for notational simplicity. And the effective mass parameters 
$M_0, \; M'$, and $ M''$ can be generated by
the  VEVs of the $SO(10)$ singlet chiral superfields.

The mass scales are assumed to satisfy  $M_0 \sim M' \sim  M'' \gg
\langle \phi^2 \rangle \sim \langle S^{2\, 2} \rangle \gg \langle A^{1\, 2}
\rangle $ where $M_0$ may be at the order of GUT scale.
And the small parameters in Yukawa matrices, 
which are used to explain the fermion masses and mixings, 
are $\epsilon \sim \langle S^{2 2} \rangle \langle 45 \rangle/M_0^2$ and
$\epsilon^\prime \sim \langle A^{1 2} \rangle /M_0$.
  In this model, the second generation masses are of order $\epsilon$, while 
the first generation masses are of order $\epsilon'^2/\epsilon$.

In the effective theory below $M_0$, after we integrate
out the Froggatt-Nielsen states  \{$\psi,\,
\bar \psi,\ \psi^{a'},\, \bar \psi_{a'},\  \psi_{a'},\, \bar \psi^{a'}$\},
 we obtain the effective Yukawa matrices for up-type
quarks, down-type quarks, and charged leptons,
 and the Dirac neutrino Yukawa matrix
\begin{eqnarray}
Y_u ~=~  \left(\begin{array}{ccc}  
\kappa_1 \, \epsilon \, \rho & (\epsilon' + \kappa_2 \, \epsilon) \rho & 0 \\
 - (\epsilon' - \kappa_2 \, \epsilon) \rho &  \epsilon \rho & \epsilon r ~T_{\bar u}     \\
      0  & \epsilon r ~T_Q& 1 \end{array} \right) \; y ~,~\,
\label{Yu}
\end{eqnarray}
\begin{eqnarray}
Y_d ~=~  \left(\begin{array}{ccc} 
 \kappa_1 \, \epsilon & \epsilon' + \kappa_2 \, 
\epsilon & 0 \\
- (\epsilon' - \kappa_2 \, \epsilon)  &  \epsilon  & \epsilon r \delta ~T_{\bar d}\\
0  & \epsilon r ~T_Q & 1 \end{array} \right) \; y ~,~\,
\label{Yd}
\end{eqnarray}
\begin{eqnarray}
Y_e ~=~  \left(\begin{array}{ccc}  
3 \, \kappa_1 \, \epsilon & - (\epsilon' - 3 \, 
\kappa_2 \, \epsilon) & 0 \\
 \epsilon' + 3 \, \kappa_2 \, \epsilon &  3 \epsilon  & \epsilon r ~T_{\bar e} \\
  0  & \epsilon r \delta ~T_L & 1 \end{array} \right) \; y ~,~\,
\label{Ye}
\end{eqnarray}
\begin{eqnarray}
Y_{\nu} ~=~  \left(\begin{array}{ccc}  
3 \, \kappa_1 \, \epsilon \, \delta' &
 - (\epsilon' - 3 \, \kappa_2 \, \epsilon) \, \delta' & 0 \\
  (\epsilon' + 3 \, \kappa_2 \, \epsilon) \,  \delta' &  3 \epsilon \delta' & 
{1 \over 2} \epsilon r \delta' ~T_{\bar \nu} \\
       0  & \epsilon r \delta ~T_L& 1 \end{array} \right) \;  y ~,~\,
\label{Ynu}
\end{eqnarray}
where
\begin{eqnarray} 
\rho ~=~ - {{5\beta_Y (1-3\alpha_{X'})} \over {16\alpha_{X'} (1+ \alpha_{X'}) }}~,~
\delta ~=~ {{1+ \alpha_{X'}}\over {1-3\alpha_{X'}}}
~,~ \delta' ~=~  {{2 \delta}\over {2 \delta-1}}~,~\,
\end{eqnarray}
\begin{eqnarray} 
T_f & = & ({\rm Baryon~ Number - Lepton~ Number}) ~,~
{\rm where}~ f = \{Q,\bar u,\bar d, L,\bar e, \bar \nu\}~.~\,
\end{eqnarray}

In the basis $\{\nu_1, \nu_2, \nu_3, {\bar \nu}_1, {\bar \nu}_2,
 {\bar \nu}_3, N_1, N_2, N_3\}$, the generalized neutrino mass matrix is 
\begin{eqnarray}
M_{\nu, {\bar \nu}, N} = \left(\matrix{ 0 & M^{\nu}_D &0 \cr
(M_D^{\nu})^T & 0 & M_{{\bar \nu}, N} \cr
0 & (M_{{\bar \nu}, N})^T & M_N \cr}\right)
~,~ \,
\label{YnuN}
\end{eqnarray}
where 
\begin{eqnarray}
(M^{\nu}_D)_{ij} &=&  (Y_{\nu})_{ij} \; {v \over\sqrt{2}}\; \sin\beta ~,~ \,
\end{eqnarray}
\begin{eqnarray}
M_{{\bar \nu}, N} ~=~ \left( \begin{array}{ccc}  3 \, \kappa_1 \, 
\epsilon \, {\tilde V}_{\overline{\psi}_H} & (\epsilon' + 3 \, \kappa_2 \, \epsilon)\, 
{\tilde V}_{\overline{\psi}_H} & 0 \\
  - (\epsilon' - 3 \, \kappa_2 \, \epsilon)\, 
{\tilde V}_{\overline{\psi}_H}   & 3 \epsilon {\tilde V}_{\overline{\psi}_H} & 0 \\
                                 0  & r \, \epsilon \, (1 - \delta) \,
T_{\bar \nu} {\tilde V}_{\overline{\psi}_H}  &
{\tilde V}'_{\overline{\psi}_H}
\end{array}\right) ~,~ \,
\end{eqnarray}
\begin{eqnarray}
  M_N ~= ~ \left( \begin{array}{ccc}  \kappa_1 \, {\tilde S} & \kappa_2 \, 
{\tilde S} & 0 \\ \kappa_2 \, {\tilde S} & {\tilde S}  & {\tilde \phi} \\
                                 0  & {\tilde \phi}  &  0 \end{array}\right) ~,~ \,
\end{eqnarray}
where ${\tilde V}_{\overline{\psi}_H}$ and ${\tilde V}'_{\overline{\psi}_H}$
are proportional to the VEV of $\overline{\psi}_H$ (with different
implicit Yukawa couplings),  ${\tilde S}$ and  ${\tilde \phi} $ 
are up to couplings the VEVs of $S^{22}$ and $\phi^2$, respectively.

It has been shown in Ref.~\cite{SRABY1} that the very good fits to 
the Standard Model fermion (quark, lepton, and neutrino) masses and mixings 
are obtained from the Yukawa matrices given in Eqs. (\ref{Yu}-\ref{Ynu})
and the generalized neutrino mass matrix given in Eq. (\ref{YnuN}).
In particular, the neutrino mixings can be bilarge, which explain
 the solar neutrino experiment, the atmospheric neutrino experiment,
and the Kamland experiment.

\subsection{Model II}
In  Model II,  the zero modes $ (\Sigma_i)^{(0)}_{{\bf (16, -3)}}$ of
 $ (\Sigma_i)_{{\bf (16, -3)}}$ where
$i=2, 3$ are considered as the first two families of fermions, and the 
 $R$ symmetry $SO(2)_{56}\times U(1)_{{\bf 4}_+}
\times SU(2)_{{\bf 4}_-}$ can give us the flavour symmetry to
 generate the fermion masses and mixings.

To be convenient in discussions, we define 
\begin{eqnarray}
{\bf 16}_1 \equiv - (\Sigma_3)^{(0)}_{{\bf (16, -3)}}
~,~ {\bf 16}_2 \equiv  (\Sigma_2)^{(0)}_{{\bf (16, -3)}}
~,~ \Upsilon_0 \equiv  (\Sigma_1)^{(0)}_{{\bf (16, -3)}}~,~\, 
\end{eqnarray}
where ${\bf 16}_i$ is the $i-th$ family of the Standard Model fermions.
In this convention, the first two families
$({\bf 16}_1, {\bf 16}_2)^T$ form a fundamental
representation under the $SU(2)_{{\bf 4}_-}$ $R$ symmetry.

We define the $U(1)_R^{II}$ and $U(1)_F^{II}$ global symmetries from 
the $SO(2)_{56} \times U(1)_{{\bf 4}_+}$ $R$ symmetry, respectively.
The generators for $U(1)_R^{II}$ and $U(1)_F^{II}$ in $SU(4)_R$ are
\begin{eqnarray}
T_{U(1)_R^{II}} ={\rm diag}\left[-1, 1, 0, 0\right]~,~\,
\label{IIU1R}
\end{eqnarray}
\begin{eqnarray}
T_{U(1)_F^{II}} ={\rm diag}\left[{1\over 2}, {1\over 2}, 
-{1\over 2}, -{1\over 2}\right]~.~\,
\end{eqnarray}
So, under $U(1)_R^{II}$, the Grassmann coordinate $\theta$ has charge
$+1$,  ${\bf 16}_1$ and ${\bf 16}_2$ have charge $+1$, while
$\Upsilon_0$ has charge 0. In addition, under $U(1)_F^{II}$,
the Grassmann coordinate $\theta$ has charge $-1/2$, 
${\bf 16}_1$ and ${\bf 16}_2$ have charge 0, while
$\Upsilon_0$ has charge $-1$.

On the observable 3-brane at $z=0$,
we introduce the third family of fermion ${\bf 16}_3$ whose quantum numbers
under $U(1)_X\times U(1)_R^{II} \times U(1)_F^{II}$ are
$(-3, +1, -1/2)$, and a ${\bf 16}$ spinor representation $\Upsilon'_0$
whose quantum numbers
under $U(1)_X\times U(1)_R^{II} \times U(1)_F^{II}$ are
$(+3, +2, 0)$. So, we can introduce the mass term 
$M_{\Upsilon_0} \Upsilon_0 \Upsilon'_0$ for
$\Upsilon_0$ and $\Upsilon'_0$ in the superpotential on the 3-brane at $z=0$.
The mass term for
$\Upsilon_0$ and $\Upsilon'_0$ can also be generated by the superpotential 
$ s_1 s'_1 \Upsilon_0 \Upsilon'_0$
where $s_1$ and $s'_1$ will be defined when we discuss the anomaly cancellations.

In Model II, the $R$ symmetry $U(1)_F^{I} \times SU(2)_{{\bf 4}_-}$
can be considered as flavour symmetry to explain the fermion masses and
mixings. Similar to the discussions in Model I, we assume that the particles,
 which are introduced to break the gauge symmetry
and generate the fermion masses and mixings, are on 
the observable 3-brane at $z=0$.
The additional particle content in Model II is the same as that
 in Model I. In short, the particles in Model II
are the Standard Model fermions ${\bf 16}_{a'}$, ${\bf 16}_{3}$;
Higgs fields $H$, ${\bf 45}_H$, ${\psi}_H$, $\overline{\psi}_H$;
vector like particles $\Upsilon_0$, $\Upsilon'_0$,
$\psi$, $\overline{\psi}$, $\psi^{a'}$, $\overline{\psi}_{a'}$,
$\psi_{a'}$, $\overline{\psi}^{a'}$; three
$SO(10)$ singlets $N_{a'}$ and $N_3$; and Higgs fields for flavour symmetry
 breaking $\phi^{a'}$, $S^{a' b'}$, and $A^{a' b'}$. 
The quantum numbers for the particles in Model II
 under the $U(1)_X$ gauge
symmetry and $U(1)_R^{II}\times U(1)_F^{II} $
$R$ symmetry are given in Table \ref{spectrumII}, 
where $Q_X$ and $Q_F$ are free parameters.\\

I. Low Energy $R$ Symmetry.\\

The $U(1)_R^{II}$ is the traditional $U(1)_R$ $R$ symmetry defined in
the low energy supersymmetry phenomenology. So, the superpartners
of the ordinary particles can be the candidates of cold dark matter,
for instance, the neutralino.\\

II. Anomaly Cancellations.\\

Similar to the discussions in the subsection 3.1, the
anomalies $[SO(10)]^2 U(1)_X$, $[U(1)_X]^3$, and $[{\rm Gravity}]^2 U(1)_X$
can be cancelled by a generalized Green-Schwarz mechanism.
Here, we discuss the anomaly cancellations by introducing
 the vector-like particles on three 3-branes at three fixed points.

Similar to the anomaly discussions in Model I,
the anomalies localized on the 3-brane at each fixed point due to the bulk
vector multiplet are equivalent to the anomalies from a multiplet
with quantum number $({\bf 16, -3})$ under the $SO(10)\times U(1)_X$
gauge symmetry. If we consider the anomaly inflows~\cite{anomaly, top},  
like the 5-dimensional orbifold theories,
the 4-dimensional anomaly cancellations are sufficient
 to ensure the consistency of higher dimensional orbifold theories~\cite{anomaly}.
Thus, we only need to consider the total anomalies due to
the bulk zero modes and the particles localized on the 3-branes at
the fixed points.

To be concrete, we still discuss 
the anomaly cancellations on the 3-brane at
each fixed point, which are similar to those in Model I.
For simplicity, we consider the case with $Q_X=0$.

 First, on the observable 3-brane at $z=0$, the anomalies due to the 
zero modes of bulk vector multiplet
are cancelled by the Higgs field $\overline{\psi}_H$. And 
to cancel the anomalies generated by the Higgs field $H$,
 we introduce one pair of 10-plets $T_{10}$  and $T_{10}'$ whose
 quantum numbers are $({\bf 10, -6})$ and $({\bf 10, 0})$ respectively
under the $SO(10)\times U(1)_X$ gauge symmetry.
In order to give the masses
to $T_{10}$  and $T_{10}'$, and keep
the $U(1)_X$  D-flatness and anomaly free on the 3-brane at $z=0$,
we introduce the
$SO(10)$ singlets $s_1$ and $s'_1$ with $U(1)_X$ charges $+6$ and $-6$,
respectively. Moreover, to forbid the mixings between the Higgs 
$H$ and $T_{10}$ or between $H$ and $T_{10}'$, we assign the following
$U(1)_R^{II}\times U(1)_F^{II}$ charges for the vector-like particles: the
$U(1)_R^{II}$ charges for $T_{10}$, $T_{10}'$, $s_1$ and $s'_1$
are 1, 1, $0$ and 0, respectively, and the $U(1)_F^{II}$ charges
for $T_{10}$, $T_{10}'$, $s_1$ and $s'_1$ are $-1/2$, $-1/2$, 0 and
$0$, respectively. In short, the 
anomalies $[SO(10)]^2 U(1)_X$, $[U(1)_X]^3$, and $[{\rm Gravity}]^2 U(1)_X$
are cancelled on the 3-brane at $z=0$. And the masses of 
$T_{10}$  and $T_{10}'$ can be generated by the superpotential
$s_1 T_{10} T_{10}'$.

Second, on the 3-brane at $z=\pi R e^{{\rm i}\pi/6}/{\sqrt 3}$,
we need to cancel the anomalies generated by a multiplet
with quantum number $({\bf 16, -3})$ under the $SO(10)\times U(1)_X$
gauge symmetry. So, we introduce a pair of the 
$\Upsilon_2$ and $\Upsilon_2'$  whose
 quantum numbers are $({\bf 16, 0})$ and $({\bf \overline{16}, +3})$ respectively
under the $SO(10)\times U(1)_X$ gauge symmetry. 
And in order to give the masses to $\Upsilon_2$  and $\Upsilon_2'$,
and keep the $U(1)_X$  D-flatness
and anomaly free on the 3-brane at $z=\pi R e^{{\rm i}\pi/6}/{\sqrt 3}$,
we introduce the
$SO(10)$ singlets $s_2$ and $s'_2$ with $U(1)_X$ charges $-3$ and $+3$,
respectively. In addition, to forbid the mixings between the SM fermions 
${\bf 16}_j$ ($j=1, 2$) and $\Upsilon_2$, and the mass terms between
 $ {\bf 16}_j$ and $\Upsilon_2'$, we assign the following
$U(1)_R^{II}\times U(1)_F^{II}$ charges for the vector-like particles: the
$U(1)_R^{II}$ charges for $\Upsilon_2$, $\Upsilon_2'$, $s_2$ and $s'_2$
are 1, 1, $0$ and 0, respectively, and the $U(1)_F^{II}$ charges
for $\Upsilon_2$, $\Upsilon_2'$, $s_2$ and $s'_2$ are $-1/2$, $-1/2$, 0 and
$0$, respectively.
And the masses of 
 $\Upsilon_2$  and $\Upsilon_2'$ can be generated by the superpotential
$s_2 \Upsilon_2 \Upsilon_2'$.

Third,  the anomaly cancellations on the 3-brane at 
$z= 2 \pi R e^{{\rm i}\pi/6}/{\sqrt 3}$ is similar to those
on the 3-brane at $z=\pi R e^{{\rm i}\pi/6}/{\sqrt 3}$. We introduce
the particles $\Upsilon_3$, $\Upsilon_3'$, $s_3$ and $s'_3$, whose
quantum numbers are the same as those for
$\Upsilon_2$, $\Upsilon_2'$, $s_2$ and $s'_2$, respectively. \\

III. Fermion Masses and Mixings.\\

With the particle quantum numbers given in Table \ref{spectrumII},
we obtain that the superpotential in Model II is the same
as that in Model I given by Eq. (\ref{superpotential}).
Therefore, the discussions for the fermion masses and mixings are
the same as those in Model I, too. The 
 effective Yukawa matrices for up-type
quarks, down-type quarks, and charged leptons,
 and the Dirac neutrino Yukawa matrix
are given by Eqs. (\ref{Yu}-\ref{Ynu}), and the
 generalized neutrino mass matrix is given by Eq. (\ref{YnuN}).

In addition, we consider the cut-off scale ($M_*$) is close to 
the compactification scale ($1/R$) due to the
 unitarity constraints~\cite{unitarity}.
Then, comparing to the Yukawa couplings of the third family,
the volume suppression factor $1/(M_* R)$ for the Yukawa couplings of
the first two families of fermions is close to 1, {\it i.e}, at the
 order of 1 (${\cal O} (1)$).

In short, we can also have very good fits to 
the Standard Model fermion masses and mixings,
especially the neutrino mixings can be bilarge.

\section{Comments on the ${\cal N} = (1, 1)$ Supersymmetric $E_6$ Models on 
$M^4\times T^2/Z_4$ and $M^4\times T^2/Z_6$}

We are only interested in the models where there are at least
 two families of the SM fermions from the zero modes of bulk vector multiplet.
In this Section, we comment on the 
${\cal N} = (1, 1)$ supersymmetric $E_6$ models on 
$M^4\times T^2/Z_n$ in which $n=4$ and 6.

The discussions for $E_6$ symmetry breaking are similar to those in the 
Section 3. The $R_{\omega}^{E_6}$ for $E_6$ gauge group is the product of the 
$R_{\omega}^{SO(10)}$ for $SO(10)$ and $R_{\omega}^{U(1)_X}$ for
$U(1)_X$, and we choose
\begin{eqnarray}
R_{\omega}^{SO(10)} &=& 
{\rm diag}[+1, +1, +1, +1, +1, +1, +1, +1, +1, +1]~,~\,
\label{PRSO10}
\end{eqnarray}
\begin{eqnarray}
R_{\omega}^{U(1)_X}  &=& {\rm exp}\left(-{\rm i} {{2 \pi Q} \over {3 n} }\right)~,~\,
\label{PRU(1)}
\end{eqnarray}
where $n=4$ and 6 for the $T^2/Z_4$ and $T^2/Z_6$ orbifolds, respectively.

In order to have two families of the SM fermions from the zero modes of bulk 
vector multiplet, we obtain $k=0$. 
Using Eqs. (\ref{Vtrans}-\ref{S3trans}) and
(\ref{PRSO10}-\ref{PRU(1)}), we obtain the transformations of vector multiplet 
\begin{eqnarray}
  V_{{\bf (45, 0)}} (x^{\mu}, ~\omega z, ~\omega^{-1} {\bar z}) ~=~ 
 V_{{\bf (45, 0)}} (x^{\mu}, ~z, ~{\bar z})  ~,~\,
\label{PVV1}
\end{eqnarray}
\begin{eqnarray}
V_{{\bf (1, 0)}} (x^{\mu}, ~\omega z, ~\omega^{-1} {\bar z}) ~=~ 
 V_{{\bf (1, 0)}} (x^{\mu}, ~z, ~{\bar z})~,~\,
\label{PVV2}
\end{eqnarray}
\begin{eqnarray}
  V_{{\bf (16, -3)}} (x^{\mu}, ~\omega z, ~\omega^{-1} {\bar z}) ~=~ \omega
 V_{{\bf (16, -3)}} (x^{\mu}, ~z, ~{\bar z}) ~,~\,
\label{PVV3}
\end{eqnarray}
\begin{eqnarray} 
V_{{\bf (\overline{16}, 3)}} (x^{\mu}, ~\omega z, ~\omega^{-1} {\bar z}) ~=~ 
\omega^{n-1} V_{{\bf (\overline{16}, 3)}} (x^{\mu}, ~z, ~{\bar z})~,~\,
\label{PVV4}
\end{eqnarray}
\begin{eqnarray}
  (\Sigma_j)_{{\bf (45, 0)}} (x^{\mu}, ~\omega z, ~\omega^{-1} {\bar z}) ~=~ \omega^{n-1}
(\Sigma_j) _{{\bf (45, 0)}} (x^{\mu}, ~z, ~{\bar z}) ~,~\,  
\label{PSS1}
\end{eqnarray}
\begin{eqnarray}
(\Sigma_j)_{{\bf (1, 0)}} (x^{\mu}, ~\omega z, ~\omega^{-1} {\bar z}) ~=~ \omega^{n-1}
  (\Sigma_j)_{{\bf (1, 0)}} (x^{\mu}, ~z, ~{\bar z})~,~\,
\label{PSS2}
\end{eqnarray}
\begin{eqnarray}
  (\Sigma_j)_{{\bf (16, -3)}} (x^{\mu}, ~\omega z, ~\omega^{-1} {\bar z}) ~=~ 
 (\Sigma_j)_{{\bf (16, -3)}} (x^{\mu}, ~z, ~{\bar z}) ~,~\, 
\label{PSS3}
\end{eqnarray}
\begin{eqnarray}
(\Sigma_j)_{{\bf (\overline{16}, 3)}} (x^{\mu}, ~\omega z, ~\omega^{-1} {\bar z}) ~=~ 
\omega^{n-2} (\Sigma_j)_{{\bf (\overline{16}, 3)}} (x^{\mu}, ~z, ~ {\bar z})~,~\,
\label{PSS4}
\end{eqnarray}
\begin{eqnarray}
  (\Sigma_3)_{{\bf (45, 0)}} (x^{\mu}, ~\omega z, ~\omega^{-1} {\bar z}) ~=~ \omega^2
(\Sigma_3) _{{\bf (45, 0)}} (x^{\mu}, ~z, ~{\bar z}) ~,~\,  
\label{PPSS1}
\end{eqnarray}
\begin{eqnarray}
(\Sigma_3)_{{\bf (1, 0)}} (x^{\mu}, ~\omega z, ~\omega^{-1} {\bar z}) ~=~ \omega^2
  (\Sigma_3)_{{\bf (1, 0)}} (x^{\mu}, ~z, ~{\bar z})~,~\,
\label{PPSS2}
\end{eqnarray}
\begin{eqnarray}
  (\Sigma_3)_{{\bf (16, -3)}} (x^{\mu}, ~\omega z, ~\omega^{-1} {\bar z}) ~=~ \omega^3
 (\Sigma_3)_{{\bf (16, -3)}} (x^{\mu}, ~z, ~{\bar z}) ~,~\, 
\label{PPSS3}
\end{eqnarray}
\begin{eqnarray}
(\Sigma_3)_{{\bf (\overline{16}, 3)}} (x^{\mu}, ~\omega z, ~\omega^{-1} {\bar z}) ~=~ 
\omega (\Sigma_3)_{{\bf (\overline{16}, 3)}} (x^{\mu}, ~z, ~ {\bar z})~,~\,
\label{PPSS4}
\end{eqnarray}
where $j=1, 2$, and $\omega=e^{{\rm i} 2\pi/n}$ ($n=4, 6$).

From the general discussions on gauge symmetry breaking by discrete
symmetry~\cite{JUNII}, we obtain that only the fields
$V_{{\bf (45, 0)}}$, $V_{{\bf (1, 0)}}$, 
 $(\Sigma_1)_{{\bf (16, -3)}}$  and $(\Sigma_2)_{{\bf (16, -3)}}$ 
have zero modes. So, the
$E_6$ gauge symmetry
is broken down to $SO(10)\times U(1)_X$ for the zero modes
by orbifold projection. 
And there are two $SO(10)$ spinor representations ${\bf 16}$ from the
zero modes of bulk vector multiplet, which can be considered
as two families of fermions. The $R$ symmetry is $SO(2)_{56}\times U(1)_{{\bf 4}_+}
\times U(1)_{{\bf 4}_-}$, which can be the origin of
flavour symmetry.

We define the zero modes $ (\Sigma_1)^{(0)}_{{\bf (16, -3)}}$ and
$ (\Sigma_2)^{(0)}_{{\bf (16, -3)}}$
of $(\Sigma_1)_{{\bf (16, -3)}}$  and $(\Sigma_2)_{{\bf (16, -3)}}$
as the first two families of fermions
\begin{eqnarray}
{\bf 16}_1 \equiv  (\Sigma_1)^{(0)}_{{\bf (16, -3)}}~,~
 {\bf 16}_2 \equiv  (\Sigma_2)^{(0)}_{{\bf (16, -3)}}~.~\, 
\end{eqnarray}

To be convenient,
 we define three $U(1)_R^{III}$,  $U(1)_F^{III}$ 
and $U(1)_{RF}$ global symmetries from 
the $R$ symmetry $SO(2)_{56}\times U(1)_{{\bf 4}_+}
\times U(1)_{{\bf 4}_-}$, whose generators in
$SU(4)_R$ are
\begin{eqnarray}
T_{U(1)_R^{III}} ~=~{\rm diag}\left[-1, 0, 0, 
+1\right]~,~\,
\label{U1R3}
\end{eqnarray}
\begin{eqnarray}
T_{U(1)_F^{III}} ~=~{\rm diag}\left[0, {1\over 2}, -{1\over 2}, 0\right]~,~\,
\end{eqnarray}
\begin{eqnarray}
T_{U(1)_{RF}} ~=~ {\rm diag}\left[{1\over 2}, - {1\over 2}, - {1\over 2},
{1\over 2}\right]~.~\,
\end{eqnarray}
Under $U(1)_R^{III}$ symmetry, the Grassmann coordinate $\theta$, 
${\bf 16}_1$ and ${\bf 16}_2$ have charge $+1$,
so, the $U(1)_R^{III}$ is the traditional $U(1)_R$ $R$ symmetry in the
low energy supersymmetry phenomenology.
And under $U(1)_F^{III}$ global symmetry, 
the Grassmann coordinate $\theta$ has charge 0,
${\bf 16}_1$ and ${\bf 16}_2$ have charges $-1/2$ and $+1/2$, respectively.
In addition, under $U(1)_{RF}$ global symmetry, 
the Grassmann coordinate $\theta$ has charge $-1/2$, and the
${\bf 16}_1$ and ${\bf 16}_2$ have charge 0.
The $U(1)_F^{III}\times U(1)_{RF}$ $R$ symmetry can be considered
as flavour symmetry.

Similar to the discussions in Section 3, the anomalies 
$[SO(10)]^2 U(1)_X$, $[U(1)_X]^3$, and $[{\rm Gravity}]^2 U(1)_X$,
can be cancelled by a generalized Green-Schwarz mechanism, 
or by adding the suitable vector-like particles on the 3-branes at the fixed points.
If we introduce  
the suitable vector-like particles on the 3-branes at the fixed points to cancel
the anomalies, it is easier if we consider the anomaly inflows among the
3-branes at the fixed points.

The supersymmetric $SO(10)$ models with $U(1)$ and extra discrete 
flavour symmetry~\cite{CASB, KBWP}
or $U(1)_A$  symmetry~\cite{NMU1} have been studied previously, and those discussions
on fermion masses and mixings may be generalized to these models. However,
how to explain the fermion masses and mixings in the supersymmetric
$SO(10)\times U(1)_X$ models with 
$U(1)_F^{III}\times U(1)_{RF}$ flavour symmetry
is still an interesting question and deserves further detail study. 

\section{Comments on the
${\cal N} = (1, 1)$ Supersymmetric $SU(9)$ and $SU(8)$ Models on 
$T^2$ Orbifolds}

In this Section,
we briefly discuss the $SU(9)$ models on the space-time $M^4\times T^2/Z_3$
where there are three or two families of fermions
 from the zero modes of bulk vector multiplet.
We also briefly discuss the $SU(8)$ Models on the space-times $M^4\times T^2/Z_3$,
$M^4\times T^2/Z_4$ and $M^4\times T^2/Z_6$ where there are three or two
 families from the zero modes of bulk vector multiplet.

\subsection{ The $SU(9)$ Models on $M^4\times T^2/Z_3$}
These models have been briefly commented in Ref.~\cite{NAHGW}.
In order to break the $SU(9)$ gauge symmetry,
we choose $R_{\omega}$ in the adjoint representation of $SU(9)$
\begin{eqnarray}
R_{\omega}^{SU(9)} &=& 
{\rm diag}[+1, +1, +1, \omega^2, \omega^2, \omega^2, 
\omega, \omega, \omega]~,~\,
\label{SU9}
\end{eqnarray}
where $\omega=e^{{\rm i} 2\pi/3}$.

Under $R_{\omega}^{SU(9)}$, the $SU(9)$ gauge symmetry is broken down to the
$SU(3)_C \times SU(3)_L \times SU(3)_R \times U(1) \times U(1)'$ gauge symmetry. 
And the generators for $U(1)$ and $U(1)'$ are
\begin{eqnarray}
T_{U(1)} ~=~{\rm diag}\left[+1, +1, +1, -1, -1, -1, 0, 0,  
0\right]~,~\,
\label{SU9A}
\end{eqnarray}
\begin{eqnarray}
T_{U(1)'} ~=~{\rm diag}\left[+1, +1, +1, +1, +1, +1, -2, -2,  
-2\right]~.~\,
\label{SU9B}
\end{eqnarray}

Under the
$SU(3)_C\times SU(3)_L\times SU(3)_R\times U(1) \times U(1)'$ gauge symmetry,
the $SU(9)$ adjoint representation ${\bf 80}$ is decomposed as
\begin{equation}
\mathbf{80} = MG(\mathbf{78}) + \mathbf{(1,1,1)}_{(0,0)} + \mathbf{(1,1,1)}_{(0,0)},
\end{equation}
where
\begin{equation}
\mathbf{MG}(\mathbf{78})~ =~ \left(
\begin{array}{ccc}
\mathbf{(8,1,1)}_{(0,0)} & \mathbf{(3, \bar 3, 1)}_{(2,0)} & 
\mathbf{(3,1,\bar 3)}_{(1,3)} \\
\mathbf{(\bar 3, 3,1)}_{(-2,0)} & 
\mathbf{(1,8,1)}_{(0,0)} & \mathbf{(1, 3, \bar 3)}_{(-1,3)} \\
\mathbf{(\bar 3,1,3)}_{(-1,-3)} & 
\mathbf{(1,\bar 3,3)}_{(1,-3)} & \mathbf{(1,1,8)}_{(0,0)}
\end{array}
\right)  ~,~\,
\end{equation}
where the subscripts denote the charges under the $U(1) \times U(1)'$
gauge symmetry.

Note that $k=0$ for $T^2/Z_3$ orbifold.
Using Eqs. (\ref{Vtrans}-\ref{S3trans}) and
(\ref{SU9}), we obtain that
the $Z_3$ transformation properties for these decomposed
 representations of the vector multiplet
$V$ and chiral multiplets $\Sigma_i$  are
\begin{equation}
V : \left(
\begin{array}{ccc}
1 & \omega & \omega^2 \\
\omega^2 & 1 & \omega \\
\omega & \omega^2 & 1
\end{array}
\right) + (1) + (1)~,~
\quad
\Sigma_i: \left(
\begin{array}{ccc}
\omega^2 & 1 & \omega \\
\omega & \omega^2 & 1 \\
1 & \omega & \omega^2
\end{array}
\right) + (\omega^2) + (\omega^2)~.~\,
\label{trans-law}
\end{equation}
So, the $\mathbf{(3, \bar 3, 1)}_{(2,0)}$, $\mathbf{(1, 3, \bar 3)}_{(-1,3)}$,
and $\mathbf{(\bar 3, 1, 3)}_{(-1,-3)}$ in $\Sigma_i$ ($i=1, 2, 3$) have
the zero modes. And these zero modes
can be considered as three families or two families
 of fermions in the ``trinification''.

Similar to the discussions in Section 3,
we can obtain the $SU(9)$ model with
$U(1)_F^I \times SU(2)_{{\bf 4}_-}$ flavour symmetry and three families of
fermions from the zero modes of bulk vector multiplet,
 and the $SU(9)$ model with
$U(1)_F^{II} \times SU(2)_{{\bf 4}_-}$
 flavour symmetry and two families from the zero modes of bulk vector multiplet. 
Therefore, how to explain the fermion masses and mixings in the extra 
``trinification'' $SU(3)_C \times SU(3)_L \times SU(3)_R \times U(1) \times U(1)'$
  with these flavour symmetries is another interesting question.
By the way, the anomalies due to the $U(1) \times U(1)'$ gauge symmetry
can be cancelled similarly.

\subsection{ The $SU(8)$ Models on $M^4\times T^2/Z_n$}

The $SU(8)$ model on $M^4\times T^2/Z_6$ has been discussed previously
where the electroweak (EW) breaking
 Higgs fields as well as the third family of fermions are
unified into the 6-dimensional vector multiplet~\cite{GMSN}.
Here, we concentrate on the models with three or two families of
fermions from the zero modes of bulk vector multiplet.\\

I. $SU(8)$ Models on $M^4\times T^2/Z_3$.\\

In order to break the $SU(8)$ gauge symmetry,
we choose $R_{\omega}$ in the adjoint representation of $SU(8)$
\begin{eqnarray}
R_{\omega}^{SU(8)} &=& 
{\rm diag}[+1, +1, +1, +1, \omega^2, \omega^2,  
\omega, \omega]~,~\,
\label{SU8I}
\end{eqnarray}
where $\omega=e^{{\rm i} 2\pi/3}$.

Under $R_{\omega}^{SU(8)}$, the $SU(8)$ gauge symmetry is broken down to the
$SU(4)_C \times SU(2)_L \times SU(2)_R \times U(1) \times U(1)'$ gauge symmetry. 
And the generators for $U(1)$ and $U(1)'$ are
\begin{eqnarray}
T_{U(1)} ~=~{\rm diag}\left[+1, +1, +1, +1, -1, -1, -1, -1\right]~,~\,
\label{SU8AI}
\end{eqnarray}
\begin{eqnarray}
T_{U(1)'} ~=~{\rm diag}\left[0, 0, 0, 0, +1, +1, -1, -1\right]~.~\,
\label{SU8BI}
\end{eqnarray}

Under the $SU(4)_C\times SU(2)_L\times SU(2)_R\times U(1) \times U(1)'$ 
gauge symmetry, the $SU(8)$ adjoint representation ${\bf 63}$ is decomposed as
\begin{equation}
\mathbf{63} = MG(\mathbf{61}) + \mathbf{(1,1,1)}_{(0,0)} + \mathbf{(1,1,1)}_{(0,0)},
\end{equation}
where
\begin{equation}
\mathbf{MG}(\mathbf{61})~ =~ \left(
\begin{array}{ccc}
\mathbf{(15,1,1)}_{(0,0)} & \mathbf{(4, 2 , 1)}_{(2,-1)} & 
\mathbf{(4,1, 2)}_{(2,1)} \\
\mathbf{(\bar 4, 2,1)}_{(-2,1)} & 
\mathbf{(1,3,1)}_{(0,0)} & \mathbf{(1, 2, 2)}_{(0,2)} \\
\mathbf{(\bar 4,1,2)}_{(-2,-1)} & 
\mathbf{(1, 2, 2)}_{(0,-2)} & \mathbf{(1,1,3)}_{(0,0)}
\end{array}
\right)  ~,~\,
\end{equation}
where the subscripts denote the charges under the $U(1) \times U(1)'$
gauge symmetry.

Note that $k=0$ for $T^2/Z_3$ orbifold.
Using Eqs. (\ref{Vtrans}-\ref{S3trans}) and
(\ref{SU8I}), we obtain that
the $Z_3$ transformation properties for these decomposed
 representations of the vector multiplet
$V$ and chiral multiplets $\Sigma_i$ are
\begin{equation}
V : \left(
\begin{array}{ccc}
1 & \omega & \omega^2 \\
\omega^2 & 1 & \omega \\
\omega & \omega^2 & 1
\end{array}
\right) + (1) + (1)~,~
\quad
\Sigma_i: \left(
\begin{array}{ccc}
\omega^2 & 1 & \omega \\
\omega & \omega^2 & 1 \\
1 & \omega & \omega^2
\end{array}
\right) + (\omega^2) + (\omega^2)~.~\,
\label{trans-law8I}
\end{equation}
So, the $\mathbf{(4, 2 , 1)}_{(2,-1)}$, $\mathbf{(\bar 4,1,2)}_{(-2,-1)}$
and $\mathbf{(1, 2, 2)}_{(0,2)}$ in $\Sigma_i$ ($i=1, 2, 3$) have
the zero modes. The zero modes of $\mathbf{(4, 2 , 1)}_{(2,-1)}$
and $\mathbf{(\bar 4,1,2)}_{(-2,-1)}$ in $\Sigma_i$ can be
considered as three families of the SM fermions in the
Pati-Salam model, while the $\mathbf{(1, 2, 2)}_{(0,2)}$ in $\Sigma_i$
can be considered as three bi-doublet EW breaking Higgs particles.
Unfortunately, these Higgs fields can not give the correct masses and mixings
for the Standard Model fermions unless we add various
Yukawa terms on the 3-branes at the fixed points, 
so, we do not consider them as the EW breaking
Higgs particles, {\it i.e.}, we give up the gauge-Higgs-fermion
unification or the gauge-Yukawa unification. For simplicity,
we introduce the Standard Model EW breaking Higgs fields
on the observable 3-brane
at the fixed point $z=0$, which are similar to those in Section 3.

Similar to the discussions in Section 3,
we can have the $SU(8)$ model with 
$U(1)_F^I \times SU(2)_{{\bf 4}_-}$ flavour symmetry
and three families of fermions from the zero modes of bulk vector multiplet,
 and the $SU(8)$ model with
$U(1)_F^{II} \times SU(2)_{{\bf 4}_-}$ flavour symmetry
and two families of fermions from the zero modes of bulk vector multiplet.
Essentially speaking, the $SO(10)$ models are
 similar to the Pati-Salam models when we consider the fermion masses
and mixings. Thus, similar to the discussions in Section 3,
we can explain the fermion masses and mixings via these flavour symmetries.
And three bi-doublets from the zero modes of bulk vector multiplet
can obtain the heavy masses if we couple them with the singlets
which have non-zero VEVs on the 3-branes at the fixed points.
By the way, the discussions for anomaly cancellations
due to the $U(1)\times U(1)'$ gauge symmetry are similar to
those in Section 3, too.\\

II. $SU(8)$ Models on $M^4\times T^2/Z_4$ and  $M^4\times T^2/Z_6$.\\

We consider the $SU(8)$ Models on the space-time
$M^4\times T^2/Z_n$ where $n=4, 6$.
In order to break the $SU(8)$ gauge symmetry,
we choose 
\begin{eqnarray}
R_{\omega}^{SU(8)} &=& 
{\rm diag}[+1, +1, +1, +1, \omega^{n-1}, \omega^{n-1},  
\omega, \omega]~,~\,
\label{SU8II}
\end{eqnarray}
where $\omega=e^{{\rm i} 2\pi/n}$ with $n=4, 6$.
Under $R_{\omega}^{SU(8)}$, the $SU(8)$ gauge symmetry is broken down to the
$SU(4)_C \times SU(2)_L \times SU(2)_R \times U(1) \times U(1)'$ gauge symmetry. 

Using Eqs. (\ref{Vtrans}-\ref{S3trans}) with $k=0$ and Eq. 
(\ref{SU8II}), we obtain that
the $Z_n$ transformation properties for these decomposed
 representations of the vector multiplet
$V$ and chiral multiplets $\Sigma_i$ are
\begin{equation}
V : \left(
\begin{array}{ccc}
1 & \omega & \omega^{n-1} \\
\omega^{n-1} & 1 & \omega^{n-2} \\
\omega & \omega^2 & 1
\end{array}
\right) + (1) + (1)~,~\quad
\Sigma_3: \left(
\begin{array}{ccc}
\omega^2 & \omega^3 & \omega \\
\omega & \omega^2 & 1 \\
\omega^3 & \omega^4 & \omega^2
\end{array}
\right) + (\omega^{2}) + (\omega^{2})~,~\,
\label{trans-II}
\end{equation}
\begin{equation}
\Sigma_j: \left(
\begin{array}{ccc}
\omega^{n-1} & 1 & \omega^{n-2} \\
\omega^{n-2} & \omega^{n-1} & \omega^{n-3} \\
1 & \omega & \omega^{n-1}
\end{array}
\right) + (\omega^{n-1}) + (\omega^{n-1})~,~\,
\label{trans-law8II}
\end{equation}
where $j=1, 2$.
So, the $\mathbf{(4, 2 , 1)}_{(2,-1)}$ and
 $\mathbf{(\bar 4,1,2)}_{(-2,-1)}$ in $\Sigma_j$ ($j=1, 2$) have
 zero modes, which can  be
considered as two families of fermions in the
Pati-Salam model. 
The $\mathbf{(1, 2, 2)}_{(0, 2)}$
 and $\mathbf{(1, 2, 2)}_{(0, -2)}$ in $\Sigma_3$ for $T^2/Z_4$
orbifold, and the  $\mathbf{(1, 2, 2)}_{(0, 2)}$ in $\Sigma_3$ for $T^2/Z_6$
orbifold have zero modes, too. 
The $\mathbf{(1, 2, 2)}_{(0, 2)}$ in $\Sigma_3$
can be considered as a bi-doublet Higgs particle.
Similar to the $SU(8)$ models on $M^4\times T^2/Z_3$, 
this bi-doublet Higgs particle can not give the correct 
Standard Model fermion masses and mixings
 unless we add various
Yukawa terms on the 3-branes at the fixed points,
 so, we do not consider it as the EW breaking
Higgs particle.

Similar to the discussions in Section 4,
we can have 
$U(1)_F^{III} \times U(1)_{RF}$ $R$ symmetry which can be considered
as flavour symmetry. Because the $SO(10)$ models
are similar to the Pati-Salam models when we consider the fermion masses
and mixings, 
like the models in Section 4, how to explain the fermion masses and
 mixings by this flavour symmetry deserves further study in detail.

\section{Discussions and Conclusions}

The other interesting flavour symmetry is $SU(3)$. There are two
possibilities for $SU(3)$ flavour symmetry: one is the gauged
flavour symmetry, the other is the global flavour symmetry.
For the gauged $SU(3)$ flavour symmetry, one can consider the
GUT group with rank 7 or higher, for example $E_8$~\cite{Adler}.
And for the global $SU(3)$ flavour symmetry, we should
consider the 10-dimensional space-time if this 
$SU(3)$ flavour symmetry arises from the $R$ symmetry.
In the 10-dimensional ${\cal N}=1$ supersymmetry, 
the $SO(6)$ ($SO(9, 1) \longrightarrow SO(3,1) \times SO(6)$)
rotation on the extra 6-dimensional space manifold
becomes the $R$ symmetry after compactification. Note that
the Lie algebra of $SO(6)$ is isomorphic to that
of $SU(4)$ and we only keep the 4-dimensional ${\cal N} =1$
supersymmetry after compactification, similar to
the discussions in Sections 2 and 3, the $SU(4)_R/U(1)_R$ or 
$SU(3)_R$ $R$ symmetry may become the flavour symmetry if we
consider the $T^6/Z_3$ orbifold. However, the 10-dimensional
${\cal N}=1$ supersymmetric
non-abelian gauge theory generically has gauge anomaly,
and can be cancelled if and only if the gauge groups are
$SO(32)$, $E_8\times E_8$, $E_8\times U(1)^{248}$ and
$U(1)^{496}$. Thus, it is difficult to
construct the GUT models with global $SU(3)$ flavour symmetry
from $R$ symmetry
because it is not easy to break these GUT groups down
to the Standard Model gauge group.

In this paper, we discuss the 6-dimensional ${\cal N} = (1, 1)$
 supersymmetric gauge theory on the space-time $M^4\times T^2$, where
the $R$ symmetry is
$SO(2)_{56}\times SU(2)_{\bf 4_+}\times SU(2)_{\bf 4_-}$.
In order to preserve only the 4-dimensional ${\cal N} = 1$ 
supersymmetry, we consider the supersymmetry breaking and $R$ symmetry breaking
on the $T^2/Z_n$ orbifolds. We find
 that for $T^2/Z_3$ orbifold, the only possibility is $k=0$ and the
$R$ symmetry is $SO(2)_{56}\times U(1)_{{\bf 4}_+}
\times SU(2)_{{\bf 4}_-}$. For $T^2/Z_4$ orbifold, the only possibility is
$k=0$, too. And for $T^2/Z_6$ orbifold, there are 
two possibilities: $k=0$ and $k=1$. The $R$ symmetry for the $T^2/Z_4$
and $T^2/Z_6$ orbifolds is $SO(2)_{56}\times U(1)_{{\bf 4}_+} 
\times  U(1)_{{\bf 4}_-}$.
We also present the transformation properties of the bulk vector multiplet under
these $R$ symmetries.

In the 6-dimensional ${\cal N} = (1, 1)$ supersymmetric gauge theory,
the Standard Model fermions may come from the zero modes of
bulk vector multiplet due to the orbifold gauge symmetry breaking, which
may explain why there are three families of fermions in the 
Standard Model. And the $R$ symmetry can give us the flavour symmetry to
explain the fermion masses and mixings. 
To be concrete, we discuss the
${\cal N} = (1, 1)$ supersymmetric $E_6$ models
 on the space-time $M^4\times T^2/Z_3$ where the gauge symmetry
$E_6$ is broken down to $SO(10)\times U(1)_X$ by orbifold projection. 
There are three $SO(10)$ spinor representation ${\bf 16}$s from the
zero modes of bulk vector multiplet, which can be considered
as three families of fermions. We study two models in detail.
In Model I, three families of the Standard Model fermions arise
from the zero modes of bulk vector multiplet, and the 
$R$ symmetry $U(1)_F^{I} \times SU(2)_{{\bf 4}_-}$
can be considered as flavour symmetry.
However, the $U(1)_R^{I}$ $R$ symmetry is not
the traditional $U(1)_R$
 $R$ symmetry in the low energy supersymmetry phenomenology,
and then, there are no cold dark matter candidates (for example neutralino)
from the superpartners. In Model II, the first two families of the 
Standard Model fermions arise
from the zero modes of bulk vector multiplet, and the 
$R$ symmetry $U(1)_F^{II} \times SU(2)_{{\bf 4}_-}$
can be considered as flavour symmetry.
 The $U(1)_R^{II}$ $R$ symmetry is the
traditional $U(1)_R$ $R$ symmetry in the low energy supersymmetry phenomenology.
In these models, the anomalies can be cancelled, and
 we have very good fits to 
the Standard Model fermion masses and mixings.

Furthermore, we comment on the
${\cal N}=(1, 1)$ supersymmetric $E_6$ models on
 $M^4\times T^2/Z_4$ and $M^4\times T^2/Z_6$ where there
are two families of the Standard Model fermions from the zero modes
of bulk vector multiplet. We also comment on the
${\cal N}=(1, 1)$ supersymmetric $SU(9)$ models on $M^4\times T^2/Z_3$,
and the $SU(8)$ models on $M^4\times T^2/Z_n$ orbifolds in which $n=3, 4, 6$.

\section*{Acknowledgments}
We would like to thank S. Adler
and S. Nandi for helpful discussions.
 This research was supported by the National
 Science Foundation under Grant No.~PHY-0070928.

\newpage

\renewcommand{\arraystretch}{1.4}
\begin{table}[t]
\caption{Model I. Under the $U(1)_X$ gauge
symmetry and $U(1)_R^{I}\times U(1)_F^{I}$
$R$ symmetry, the quantum numbers for 
the Standard Model fermions ${\bf 16}_{a'}$, ${\bf 16}_{3}$;
Higgs fields $H$, ${\bf 45}_H$, ${\psi}_H$, $\overline{\psi}_H$;
vector like particles
$\psi$, $\overline{\psi}$, $\psi^{a'}$, $\overline{\psi}_{a'}$,
$\psi_{a'}$, $\overline{\psi}^{a'}$; three
$SO(10)$ singlets $N_{a'}$ and $N_3$; and Higgs fields for flavour symmetry
breaking $\phi^{a'}$, $S^{a' b'}$, and $A^{a' b'}$. 
In this Table, $Q_X$, $Q_R$ and $Q_F$ are free parameters.
And the $U(1)_R^{I}\times U(1)_F^{I}$ charges for ${\psi}_H$
($Q_{{\psi}_H}^R$ and $Q_{{\psi}_H}^F$) are arbitrary because we have no
constraints from superpotential.
\label{spectrumI}}
\vspace{0.4cm}
\begin{center}
\begin{tabular}{|c|c|}
\hline        
Particles &  $U(1)_X\times U(1)_R^{I}\times U(1)_F^{I}$  \\ 
\hline
${\bf 16}_{a'}$ & ($-3$; $-1$; $-1/2$ ) \\
\hline
${\bf 16}_{3}$ & ($-3$; $-1$; $+1$ ) \\
\hline
$H$ & ($+6$; $-1$; $-2$ ) \\
\hline
 ${\bf 45}_H$  &  ($-Q_X$; $-Q_R$; $-Q_F+3$)\\
\hline
${\psi}_H$ & ($0$; $Q_{{\psi}_H}^R$; $Q_{{\psi}_H}^F$) \\
\hline
 $\overline{\psi}_H$ & ($Q_X/2+3$; $Q_R/2-1/2$; $Q_F/2-5/2$ ) \\
\hline
$\psi$ & ($-Q_X - 3$; $-Q_R-1$; $-Q_F+4$) \\
\hline
 $\overline{\psi}$ &
($Q_X + 3$; $Q_R-2$; $Q_F-4$) \\
\hline
 $\psi^{a'}$ & ($ - 3$; $-1$; $5/2$)  \\
\hline
$\overline{\psi}_{a'}$ & ($ + 3$; $-2$; $-5/2$) \\
\hline
$\psi_{a'}$ & ($-Q_X - 3$; $-Q_R-1$; $-Q_F+5/2$) \\
\hline
$\overline{\psi}^{a'}$ & ($Q_X + 3$; $Q_R-2$; $Q_F-5/2$) \\
\hline
$N_{a'}$ & ($-Q_X/2$; $-Q_R/2-3/2$; $-Q_F/2$) \\
\hline
$N_3$ & ($-Q_X/2$; $-Q_R/2-3/2$; $-Q_F/2+3/2$) \\
\hline
 $\phi^{a'}$ & ($Q_X$; $Q_R$; $Q_F-3/2$) \\
\hline
$S^{a' b'}$ & ($Q_X$; $Q_R$; $Q_F$) \\
\hline
$A^{a' b'}$  & ($0$; $0$; $+3$)  \\
\hline
\end{tabular}
\end{center}
\end{table}

\renewcommand{\arraystretch}{1.4}
\begin{table}[t]
\caption{Model II. Under the $U(1)_X$ gauge
symmetry and $U(1)_R^{II}\times U(1)_F^{II}$
$R$ symmetry, the quantum numbers for 
the Standard Model fermions ${\bf 16}_{a'}$, ${\bf 16}_{3}$;
Higgs fields $H$, ${\bf 45}_H$, ${\psi}_H$, $\overline{\psi}_H$;
vector like particles $\Upsilon_0$, $\Upsilon'_0$,
$\psi$, $\overline{\psi}$, $\psi^{a'}$, $\overline{\psi}_{a'}$,
$\psi_{a'}$, $\overline{\psi}^{a'}$; three
$SO(10)$ singlets $N_{a'}$ and $N_3$; and Higgs fields for flavour symmetry
breaking $\phi^{a'}$, $S^{a' b'}$, and $A^{a' b'}$. 
In this Table, $Q_X$ and $Q_F$ are free parameters.
And the $ U(1)_F^{II}$ charge for ${\psi}_H$
( $Q_{{\psi}_H}^F$) is arbitrary because we have no
constraints from superpotential.
\label{spectrumII}}
\vspace{0.4cm}
\begin{center}
\begin{tabular}{|c|c|}
\hline        
Particles &  $U(1)_X\times U(1)_R^{II}\times U(1)_F^{II}$  \\ 
\hline
${\bf 16}_{a'}$ & ($-3$; $+1$; $0$ ) \\
\hline
${\bf 16}_{3}$ & ($-3$; $+1$; $-1/2$ ) \\
\hline
$H$ & ($+6$; $0$; $0$ ) \\
\hline
 ${\bf 45}_H$  &  ($-Q_X$; $0$; $-Q_F-1$)\\
\hline
${\psi}_H$ & ($0$; $0$; $Q_{{\psi}_H}^F$) \\
\hline
 $\overline{\psi}_H$ & ($Q_X/2+3$; $0$; $Q_F/2+1/2$ ) \\
\hline
$\Upsilon_0$ & ($-3$; $0$; $-1$)\\
\hline
$\Upsilon'_0$ & ($+3$; $2$; $0$)\\
\hline
$\psi$ & ($-Q_X - 3$; $+1$; $-Q_F-3/2$) \\
\hline
 $\overline{\psi}$ &
($Q_X + 3$; $+1$; $Q_F+1/2$) \\
\hline
 $\psi^{a'}$ & ($ - 3$; $+1$; $-1$)  \\
\hline
$\overline{\psi}_{a'}$ & ($ + 3$; $+1$; $0$) \\
\hline
$\psi_{a'}$ & ($-Q_X - 3$; $+1$; $-Q_F-1$) \\
\hline
$\overline{\psi}^{a'}$ & ($Q_X + 3$; $+1$; $Q_F$) \\
\hline
$N_{a'}$ & ($-Q_X/2$; $+1$; $-Q_F/2-1/2$) \\
\hline
$N_3$ & ($-Q_X/2$; $+1$; $-Q_F/2-1$) \\
\hline
 $\phi^{a'}$ & ($Q_X$; $0$; $Q_F+1/2$) \\
\hline
$S^{a' b'}$ & ($Q_X$; $0$; $Q_F$) \\
\hline
$A^{a' b'}$  & ($0$; $0$; $-1$)  \\
\hline
\end{tabular}
\end{center}
\end{table}

\end{document}